\documentclass[a4paper,11pt]{article}
\usepackage{amsthm,amsfonts, indentfirst, latexsym, bm,graphicx,amsmath,amssymb,mathrsfs,verbatim,hyperref,pdfpages}
\usepackage{caption}
\usepackage{subcaption}
\usepackage{color}
\usepackage[T1]{fontenc}
\usepackage{dsfont}
\usepackage{latexsym}
\usepackage{url}
\usepackage{array}
\usepackage{tikz}
\usepackage{pgfplots}
\usepackage{graphicx}
\usepackage[titletoc,title]{appendix}
\usepackage{bbm} 
\usepackage{algorithm,algorithmic}
\numberwithin{equation}{section}

\usepackage{chngcntr}
\counterwithout{equation}{section} 
\usetikzlibrary{arrows, automata,positioning,calc,shapes,decorations.pathreplacing}
  \colorlet{greencolor}{green!50!black}
  \colorlet{textcolor}{red}
  \colorlet{tancolor}{orange!80!black}
  \colorlet{bluecolor}{blue}
\title{Estimation of Semi-Markov Multi-state Models:
A Comparison of the Sojourn Times and
Transition Intensities Approaches}

\author{Azam Asanjarani\footnote{The University of Auckland.}, 
Benoit Liquet \footnote{Macquarie University \& Universit\'{e} de Pau et des Pays de \l'Adour.},
Yoni Nazarathy\footnote{The University of Queensland.}.}

\date{ }
\begin{document}
\maketitle
\begin{abstract}

Semi-Markov models are widely used for survival analysis and reliability analysis. In general, there are two competing parameterizations and each entails its own interpretation and inference properties. On the one hand, a semi-Markov process can be defined based on the distribution of sojourn times, often via hazard rates, together with transition probabilities of an embedded Markov chain. On the other hand, intensity transition functions may be used, often referred to as the hazard rates of the semi-Markov process. We summarize and contrast these two parameterizations both from a probabilistic and an inference perspective, and we highlight relationships between the two approaches. In general, the intensity transition based approach allows the likelihood to be split into likelihoods of two-state models having fewer parameters, allowing efficient computation and usage of many survival analysis tools. {Nevertheless, in certain cases the sojourn time based approach is natural and has been exploited extensively in applications.} In contrasting the two approaches and contemporary relevant R packages used for inference, we use two real datasets highlighting the probabilistic and inference properties of each approach. This analysis is accompanied by an R vignette.

\end{abstract}
\section{Introduction \label{section intro}}

In biostatistics, many  models for survival and reliability analysis are two-state stochastic processes which lead to a particular event such as death, or the outcome of a particular drug treatment. However, applying a  multi-state stochastic process allows the modeller to provide a richer and more accurate model by  adding more details. These details allow to capture alternative paths to the event of interest, specify all the intermediate events, and also allow to understand partial failure modes in a progressive disease. 

 In our context, a multi-state stochastic process is a process $X(t)$ for $t \geq 0$, where $X(t)$ can take a finite number of values $ 1,2, \ldots ,\ell$. This process can be considered as a family of random variables $X(t)$ indexed by $t$.
The quantities of interest  are often
the probability of being in a state at a given time and the distribution of first  passage times (the time until the process reaches a given state for the first time from a particular starting state).  

In some applications of  multi-state  stochastic processes, the dependence on the history of the process is  negligible. Therefore, for the sake of mathematical tractability, assuming the Markov property (where future transitions between states depend only upon
the current state) is convenient. For instance,  continuous time Markov chains, which we refer to here as {\em Markov processes},  are widely used in modelling the movements of patients between units of a hospital or between stages of a disease, see for instance \cite{broyles2010statistical, taylor1998using}, or 
in the prediction of the incidence rate of diseases, see \cite{aalen1997markov}. 

However, in certain cases, the Markov assumption is unrealistic. For instance, in the study of human sleep stages,  the sleep stages usually do not follow an exponential distribution (constant hazard rate), but can rather have other forms such as a Weibull distribution, see  \cite{wang2013computational}. 
Further, some aspects of systems' behavior can not be captured by Markov processes. For instance, the risk of chronic diseases such as  AIDS essentially depends on the time since infection, see \cite{joly1999penalized}. For these cases, applying  the class of \textit{semi-Markov processes}\index{semi-Markov process} (SMP), as an extension of Markov processes, is fruitful. Here, future probability transitions depend on the sojourn time (the time spent in the current state), and the clock of each sojourn time is reset to zero after each transition into a new state. 
SMPs have a variety of applications in healthcare. For instance,  for predicting a disease progression \cite{goshu2013modelling},  health care manpower supply prediction \cite{trivedi1987semi}, and  recovery progress of patients with a specific type of disease \cite{polesel1986application}.

For biomedical applications, especially those concerned with characterizing an individual's progression through various stages of a disease, a three-state semi-Markov process known as the \textit{illness-death model }is very popular (see for instance \cite{joly1999penalized, boucquemont2014should}).
The  illness-death model may  also be applied for modelling the trajectory of patients  in intensive care units (ICUs) of hospitals (see \cite{liquet2012investigating, coeurjolly2012attributable}). 

Here, our main focus is on  the statistical methodology of semi-Markov processes and more precisely on parametric models of SMPs.  We compare and contrast two approaches for defining SMPs which we denote via, {\bf I - sojourn times}, and {\bf II - intensity transition functions}. 

Approach I  is based on the specification of the {\em sojourn time distribution,} together with a transition probability matrix of a discrete time Markov chain, which we call the {\em embedded chain}. Approach II, is based on {\em intensity transition functions} which when multiplied by an infinitesimal quantity, specify the instantaneous probability of transition between states. Note that in the literature, these are sometimes also called hazard rate functions of the SMP, however they should not be confused with hazard functions of probability distributions (such as for example the sojourn time distributions).

While from a probabilistic perspective, both Approach I and Approach II are equivalent ways for describing an SMP, from a statistical perspective there are differences. In this paper, we highlight that when it comes to parameter estimation, Approach II has significant advantages over Approach I. Intrinsically this is because Approach II can be expressed by using fewer parameters.  Further, we can show that the likelihood function of Approach II can be written as the product of likelihoods of two-state models. This is very  helpful for reducing the  computational effort for likelihood-based parameter estimation. Nevertheless, depending on the application at hand, using either approach may be useful and in certain cases researchers have chosen to use Approach~I because it focuses directly on the underlying distributions. We highlight the associated tradeoffs in the paper. 
 
The remainder of this paper is structured as follows. In Section~\ref{section:SMP}, we introduce semi-Markov processes (SMP) via both Approach I and Approach II. We also present the probabilistic relationships between the approaches. We continue with Section~\ref{section:inference} focusing on inference and specify the likelihood function for both approaches. In Section~\ref{sec:application}, we illustrate inference on two available datasets and present how to implement both approaches (for both datasets) using several contemporary R software packages. For both datasets, we compare results of fully parametric models based on both approaches and highlight the implications of each modelling choice. Section~\ref{sec:conclusion} presents some concluding remarks. All the numerical results (tables and figures) from the paper are freely reproducible via R code in a comprehensive detailed vignette available in \cite{Liq_github}.

\section{Semi-Markov Processes}
\label{section:SMP}

\begin{figure}[h]
\center
\includegraphics[scale=0.6]{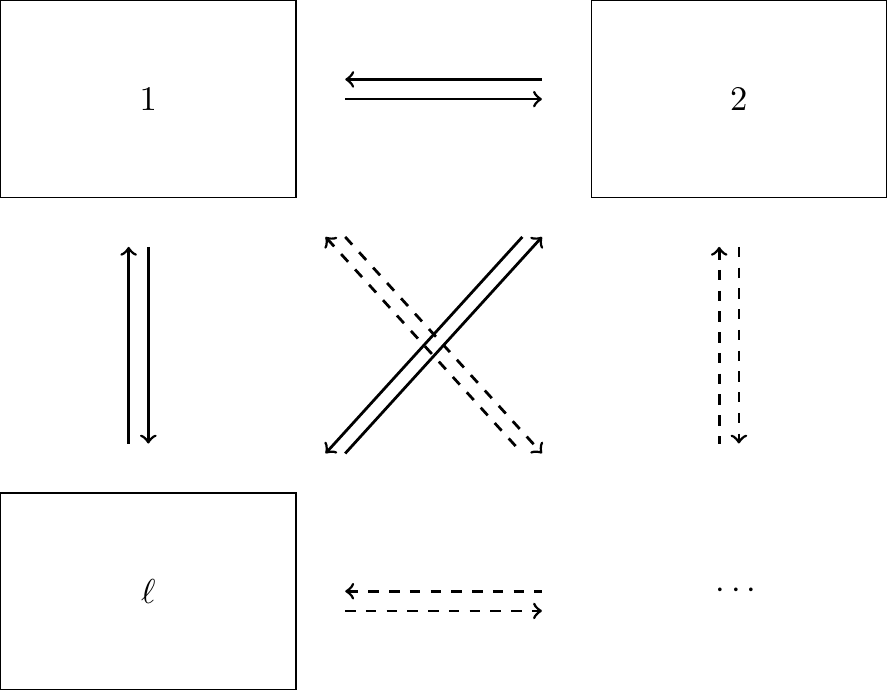}
\caption{{{\small An Illustration of a multi-state Model.  }}}
\label{Fig:Multi}
\end{figure}

A multi-state model is a continuous time stochastic process with values in a discrete set that is often applied for longitudinal medical studies, where  the patients  may experience several events and their related information is collected over time.  The complexity of a model greatly depends on the
number of states and also on the possible transitions. 
Figure~\ref{Fig:Multi} demonstrates a general multi-state model. See  \cite{andersen2012statistical}  as a  classical reference for multi-state models. 

A specific class of examples of multi-state models is the class of Markov processes where the state evolution jumps between levels and the process adheres to the Markov property. However, in many real-world applications, we need a stochastic process that exhibits dependence  between jump times. For instance, in biostatistics,  where the trajectory of  patients in a hospital  is considered, using a Markov process as a multi-state model imposes a stringent limitation on the sojourn time distribution in each state. This is because a key consequence of using Markov processes is the fact that state sojourn times follow the exponential distribution and is ``non-ageing'' with a constant hazard rate. In contrast, in Semi-Markov processes this assumption can be relaxed. Hence with Markov processes one is often lacking the desired degrees of freedom for modeling the dependence between jump times. 

Extending the modelling from Markov processes to the class of SMPs removes the restriction of memoryless (exponential) state sojourn times and at the same time, preserves the usefulness of treating the data as jump processes in continuous time. In fact, for semi-Markov processes we consider a  relaxation of the Markov property for sojourn times and only the embedded chain of states is required to follow the Markov property. For this reason, SMPs are applied for modelling a variety of phenomena in different areas such as  economics, reliability, and health care, see \cite{janssen2013semi}. 

To define an SMP, consider a  homogeneous Markov chain $\{J_n\}_{n \ge 0}$ on states $\{1,2,\ldots, \ell\}$ where the probability of $n$-th ($n\geq 1$) jump from state $i$ to state $j$ for $i \neq j$ is $p_{ij}$. That is,
\begin{equation}
\label{eq:pij}
p_{ij}=\mathbb{P}(J_{n}=j ~|~ J_{n-1}=i).
\end{equation}
Based on the directed graph associated with these probabilities (where arc $i \to j$ exists only if $p_{ij} > 0$) states can be classified as either transient or recurrent. {A state $i$ is {\em absorbing} if $p_{ij} = 0$ for all $j \neq i$. Note that under this definition 
\[
\sum_{j \neq i} p_{ij} = 
\begin{cases}
1 & \text{for non-absorbing}~i, \\
0 &  \text{for absorbing}~i.
\end{cases}
\]
}

Denote the increasing sequence of jump times by  $T_0=0< T_1< T_2 < T_3 <\ldots$  and {define}  $N(t) = \max \,\{n : T_n \leq t\}$ for $t\geq 0$. This is a count of the number of transitions up to time $t$. The  stochastic  process $X(t):= J_{N(t)}$   is said to be  a {\em semi-Markov process (SMP)} if  whenever the process enters state $i$, the next state is $j$ with probability 
$p_{ij}$
and given that the next state to be entered is $j$, the time until the transition from $i$ to $j$ is a random variable with cumulative distribution function $F_{ij}(t)$:
\begin{equation}
\label{eq:Fij}
F_{ij}(t)=\mathbb{P}(\tau_n \leq t ~|~ J_{n-1}=i, J_{n}=j), \quad t\geq 0,
\end{equation}
where $\tau_n= T_n-T_{n-1}$, see Chapter 4 of \cite{ross1996stochastic}. Hence in general, SMPs are not Markov processes as they do not posses the Markov property. Further, a semi-Markov process allows arbitrarily distributed sojourn times in any state  but retains the Markov property for the embedded (discrete time) Markov chain, $\{J_n\}_{n \ge 0}$.

In many applications of SMPs in healthcare, a very popular three state semi-Markov process known as the \textit{illness-death model} \index{illness-death model} is applied, see for instance \cite{polesel1986application}. 
%
\begin{figure}[h]
\center
\includegraphics[scale=0.6]{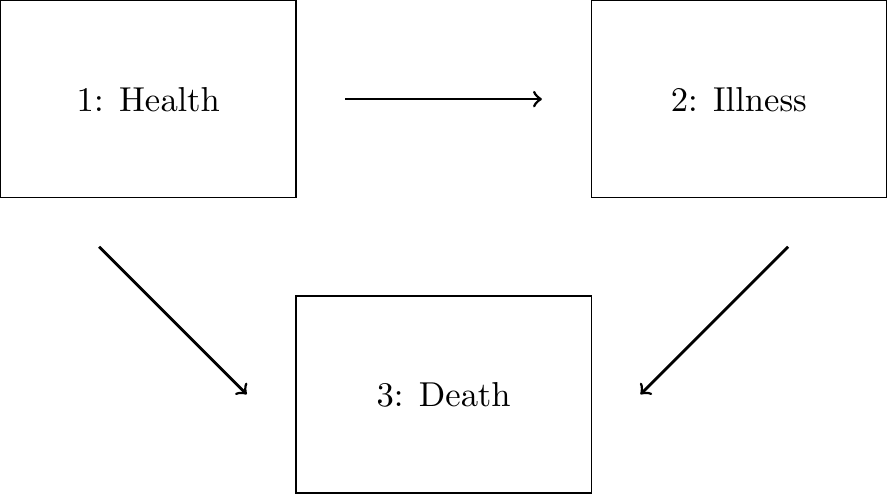}
\caption{\label{Fig:IllD}{{\small The Illness-Death Model. }}}
\end{figure}
\paragraph{Illness-death model}

\begin{figure}[h]
\center
\includegraphics[scale=0.6]{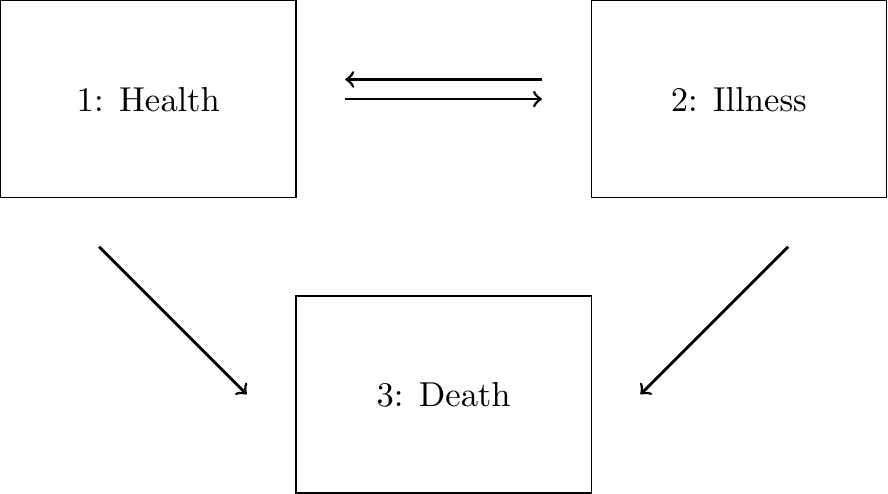}
\caption{\label{Fig:33}{{\small The Reversible Illness-Death Model. }}}
\end{figure}

The illness-death model is the most common model in epidemiology (see Figure~\ref{Fig:IllD}), where it is often applied in studying chronic diseases. 
 In this model, we have three states ``Health'', ``Illness' and ``Death', denoted by  1, 2 and 3, respectively. There are three kinds of transitions:
$1 \to 2 $, $2 \to 3 $ and $1 \to 3$ and state $3$ is absorbing. Since this model is often used to describe severe illnesses, there is no possibility of recovery, and therefore the model is irreversible\footnote{To avoid confusion, note that this term is also used for the different concept of a ``reversible'' Markov chain appearing in a different setting.}. In cases where treatments may yield remission, it is more  appropriate to construct a model with an additional state
``Remission'' rather than to consider that there is a possibility of moving back to the ``Healthy'' state.
 Here, we consider the general multi-state model with all possible transitions as illustrated  in Figure~\ref{Fig:Multi} which includes both illness-death model and its reversible version (see Figure~\ref{Fig:33}).  

We now arrive at our key point dealing with SMPs. Here we contrast the two approaches which we denote via, {\bf I - sojourn times}, and {\bf II - intensity transition functions}. Approach I was already used to define SMPs above. We spell out further details of Approach I and then continue to introduce Approach II.

\subsection{Approach  I: Sojourn Times}

As defined above, an SMP, $X(t)$, can be constructed via the sequence $\{(J_n, T_n)\}_{n\ge 0}$, of states and jump times respectively. The underlying parameters of this construction involve the transition probabilities of the embedded chain, $p_{ij}$ presented in \eqref{eq:pij}, as well as the distributions of the sojourn times for each transition $i \to j$, presented in \eqref{eq:Fij}. 

{The transition probabilities of the embedded chain are often organized in a matrix, $P = [p_{ij}]$ (where we set $p_{ii} = 0$). Note that $P$ restricted to the non-absorbing states is a stochastic matrix.} Further, assuming the sojourn time distributions are continuous, they are often represented in different forms including the probability density function, the survival function or the hazard rate function. We now present the details.

The \textit{probability density function} of the sojourn time is
\begin{equation}
\label{eq:fijI}
f_{ij}(t) = \lim_{\bigtriangleup t \rightarrow 0} \frac{1}{\bigtriangleup t}\mathbb{P}(\tau_n\in (t, t+\bigtriangleup t) ~|~J_{n-1}=i,  J_{n}=j).
\end{equation}
The corresponding  \textit{survival function} is
\begin{equation}
\label{eq:SijI}
S_{ij}(t) = \mathbb{P}(\tau_n>t ~|~ J_{n-1}=i,  J_{n}=j)=1-F_{ij}(t). 
\end{equation}
(Note that  $S_{ij}(t)$ is a decreasing function, that is $S_{ij}(0)=1$  and $\lim_{t \rightarrow +\infty} S_{ij}(t)=~0$). The \textit{hazard function} \index{hazard rate function} which is often thought of as the probability that a jump occurs in a specified interval $(t, t+\bigtriangleup t) $ given no jump before time $t$, is
\begin{equation}
\label{eq:alphaijI}
{\alpha}_{ij}(t) = \lim_{\bigtriangleup t \rightarrow 0} \frac{1}{\bigtriangleup t}\mathbb{P}\big(\tau_n \in (t, t+\bigtriangleup t)  ~|~ J_{n-1}=i,  J_n=j, \tau_n>t).
\end{equation}
 Here, note that by definition of conditional probability we have
\begin{equation}\label{eq:m}
\alpha_{ij}(t)=\frac{f_{ij}(t)}{S_{ij}(t)},
\qquad
\mbox{and}
\qquad
f_{ij}(t)=  \alpha_{ij}(t) e^{-\int_0^t \alpha_{ij}(u) du}.
\end{equation}

It is also useful to define the probability of staying in a current state $i$ for at least $t$ time units. We call this the survival function of the waiting/holding time in state $i$, and denote it via
\begin{equation}
\label{eq:S-i}
S_{i}(t)= \mathbb{P}(\tau_n>t ~|~ J_n=i) = \sum_{j \neq i} p_{ij} S_{ij}(t).
\end{equation}

\subsection{Approach II, Intensity Transition Functions}
The first approach required specification of the parameters using two types of objects, transition probabilities of the embedded Markov chain, and the distribution of sojourn times given a transition $i \to j$.  The second approach which we present now is more succinct in that it only requires one type of object: {\em intensity transition functions}. These functions are defined via
\begin{equation}
\label{eq:alphaTildeij}
 \tilde{\alpha}_{ij}(t) = \lim_{\bigtriangleup t \rightarrow 0} \frac{1}{\bigtriangleup t}\mathbb{P}( \tau_n \in (t, t+\bigtriangleup t), J_{n}=j ~|~ J_{n-1}=i, \tau_n>t),
\end{equation}
and are similar in nature to hazard rates. However, they should not be treated as hazard rate functions of a probability distribution. They rather indicate the instantaneous probability of making a transition from state $i$ to state $j$ after spending $t$ time units in state $i$ since the last transition. However summing up over all target states $j$, we do obtain a hazard rate of a probability distribution which we denote via
$
\tilde{\alpha}_i(t) = \sum_{j \neq i} \tilde{\alpha}_{ij}(t).
$
This is the hazard rate of the waiting/holding time in state $i$. We can use this approach to obtain an alternative expression to $S_i(t)$ of \eqref{eq:S-i}. 
\begin{equation}
\label{eq:SiFromAlphaTilde}
S_i(t) = e^{-\int_0^t \tilde{\alpha}_i(u) \, du} = e^{-\int_0^t \sum_{j \neq i} \tilde{\alpha}_{ij}(u) \, du}.
\end{equation}

Using Approach II, more formally, the meaning of the intensity transition functions is
\begin{equation}
\label{eq:uglyDef}
\lim_{\Delta t \to 0} \frac{\mathbb{P}\big(X(t+u + \Delta t) = j ~|~ X(t) =i, X(t^-) \neq i, \tau_{N(t)+1} > u\big)}{\Delta t} = \tilde{\alpha}_{ij}(u),
\end{equation}
where using the notation defined previously, $\tau_{N(t)+1}$ is the time elapsed since the last transition, measured at time $t$. The conditional probability in \eqref{eq:uglyDef} is conditional on the fact that the last transition was at time $t$ into state $i$ and up to time $t+u$ there have not been any further transitions. Given this condition, it indicates the instantaneous probability of: (i) making a transition at time $t+u$. (ii) Making the transition into state $j$.

It can be shown that specification of the intensity transition functions $\tilde{\alpha}_{ij}(\cdot)$ completely specify the probability law of an SMP. See \cite{breuer2005introduction} for a formal description where one can also use the notation ${\cal H}(t^-)$ to define the history of the process just before time $t$, i.e. up to $t^-$. This can formally be defined as the sigma algebra in a {filtration} associated with the stochastic process. Related notation often used in the literature is the (conditional on history) transition probability to be in state $j$ at time $t+u$, after being in state $i$ in time $t$. This is often {denoted by} $P_{ij}(t, t+u)=\mathbb{P}\big(X(t+u) = j ~|~ X(t) =i, {\cal H}(t^-)\big)$.

{We note that Approach~II can be viewed as a collection of competing risk models where at the instance of arriving to a new state, a new set of cause-specific hazard rates is defined. See for example \cite{andersen2002competing} where a competing risk model is presented as a multi-state model and the cause-specific hazard rates terminology is used.}

\subsection{Relations Between the Two Approaches}
\label{sec:relations}

We now wish to show how for the same SMP, one can use either the parameters of Approach~I or the parameters of Approach~II and interchange between them.  A key relationship is the following
\begin{equation}
\label{eq:relation-alpha-alphatilde}
\tilde{\alpha}_{ij}(t)= \frac{p_{ij}S_{ij}(t)}{S_{i}(t)} \alpha_{ij}(t) = p_{ij} \frac{f_{ij}(t)}{S_i(t)}.
\end{equation}
It is established using the conditional probabilities, defined in \eqref{eq:pij}, \eqref{eq:SijI}, \eqref{eq:S-i}, and \eqref{eq:alphaijI}. This key relationship also yields an interpretation of the {\em cumulative incidence function} for the $i \to j$ transition which we denote via $\text{CIF}_{ij}(\cdot)$. This is a common measure used in the field of competing risks, see for example \cite{koller2012comp}, that determines the probability of transitioning into $j$ before or at time $t$. By rearranging and integrating both sides of \eqref{eq:relation-alpha-alphatilde} we obtain the following representation of the cumulative incidence function.
\begin{equation}
\label{eq:cumulativeIncidence}
\text{CIF}_{ij}(t) = \int_0^t S_i(u)  \tilde{\alpha}_{ij}(u) \, du = \int_0^t f_{ij}(u)  p_{ij} \, du = p_{ij} F_{ij}(t).
\end{equation}

We can now convert between the parameterizations of both approaches as follows.

{\bf Approach I $\to$ Approach II.}  Given $\alpha_{ij}(\cdot)$ and $p_{ij}$, obtain $\tilde{\alpha}_{ij}(\cdot)$ as follows.

\begin{equation}
\label{eq:ItoII}
\tilde{\alpha}_{ij}(t)=  p_{ij} \frac{ e^{-\int_0^t  \alpha_{ij}(u) \, du}}{\sum_{k \neq i} p_{ik}  e^{-\int_0^t  \alpha_{ik}(u) \, du}}  \alpha_{ij}(t).
\end{equation}
This follows directly from \eqref{eq:relation-alpha-alphatilde}.

{\bf Approach II $\to$ Approach I.}  Given $\tilde{\alpha}_{ij}(\cdot)$ obtain $p_{ij}$ and  $\alpha_{ij}(\cdot)$:

First we have,
\begin{equation}
\label{eq:pijIItoI}
p_{ij} = \int_0^\infty \tilde{\alpha}_{ij}(t) e^{-\int_0^t \sum_{k \neq i} \tilde{\alpha}_{ik}(u) \, du} \, dt.
\end{equation}
This can also be obtained from \eqref{eq:cumulativeIncidence} by taking $t \to \infty$. Once $p_{ij}$ values are at hand we again use \eqref{eq:relation-alpha-alphatilde} and isolate $f_{ij}(t)$ to obtain,
\begin{equation}
\label{eq:densTildeTrans}
f_{ij}(t)  = \tilde{\alpha}_{ij}(t) \, \frac{e^{-\int_0^t \sum_{k \neq i} \tilde{\alpha}_{ik}(u) \, du}}{p_{ij}}, 
\end{equation}
from which we can obtain $\alpha_{ij}(t)$ using \eqref{eq:m} in the standard manner.

\subsection{Examples}
\label{sec:examples}

We now present three examples illustrating the relationship between the two approaches. 


\noindent
\paragraph{Example 1: Continuous Time Markov Chains (CTMC)}  

A (finite state) CTMC is a multi-state stochastic process on states $\{1,\ldots,\ell\}$ that is time homogenous and satisfies the Markov property. That is,
$
{\mathbb P}\big(X(s+t) = j ~|~ X(s) = i, {\cal H}(s^-)\big) = {\mathbb P}\big(X(t) = j ~|~ X(0) = i \big).
$
{For simplicity of notation we restrict attention to CTMCs without absorbing states.}
Such chains can be parameterized in several ways, two of which are reminiscent of Approach~I, and Approach~II above:
\begin{itemize}
\item CTMC Approach I: Define a $\ell \times \ell$ stochastic transition probability matrix $P$ (non-negative entries and rows sum to $1$) with $0$ entries in the diagonal. Denote the entries via $P_{ij}$. This is often called the transition probability matrix of the embedded discrete time Markov chain. Then define a vector of rates of length $\ell$, {\color{blue}$\boldsymbol {\lambda}$}, where $\lambda_i^{-1}$ denotes the mean holding time of state $i$. Then from the theory of CTMCs, see for example \cite{breuer2005introduction}, the process evolves as follows. If at time $t$ the process is in state $i$, an exponential random variable with parameter $\lambda_i$ is generated to determine a holding duration. After that duration passes a transition from state $i$ to state $j$ occurs with probability $P_{ij}$. The durations and the transition choice are independent. The process then repeats. It is now evident that such a description of a CTMC is a special case of {\bf Approach I} for SMPs, where 
\begin{equation}
\label{eq:CTMC1asSMP}
\alpha_{ij}(t) = \lambda_i
\qquad
\mbox{and}
\qquad
p_{ij} = P_{ij}.
\end{equation}
Observe that $\alpha_{ij}(t)$ is independent of $t$ and independent of the target state $j$.
\item CTMC Approach II: Define a $\ell \times \ell$ generator matrix $Q$ (non-negative entries on the off-diagonal, {negative} entries on the diagonal, and rows sum to $0$). Denote the entries via $q_{ij}$. Then treat the process as an SMP using {\bf Approach II} with 
\begin{equation}
\label{eq:CTMC2asSMP}
\tilde{\alpha}_{ij}(t) = q_{ij},
\end{equation}
Observe that $\alpha_{ij}(t)$ is independent of $t$.
\end{itemize}

It is well known from the theory of CTMCs the two parameterizations {are equivalent}.

\noindent
{\bf CTMC Approach I $\to$  CTMC Approach II}:
\[
q_{ij} = 
\begin{cases}
\lambda_i P_{ij} & i \neq j,\\
-\lambda_i & i=j, 
\end{cases}
\]
{\bf CTMC Approach II $\to$  CTMC Approach I}:
\[
P_{ij} = \frac{q_{ij}}{ \sum_{k \neq i} q_{ik}},
\qquad
\mbox{and}
\qquad
\lambda_i = \sum_{k \neq i} q_{ik} = -q_{ii}.
\]
We now verify that these transformations directly agree with the relationship between Approach I of the SMP and Approach II of the SMP. Using \eqref{eq:ItoII} and \eqref{eq:CTMC1asSMP} we have,
\begin{equation}
\label{eq:benoitGivingCourseInDenemark}
\tilde{\alpha}_{ij}(t) = P_{ij} \frac{e^{-\int_0^t \lambda_i \, du}}{\sum_{k \neq i} P_{ik} e^{-\int_0^t \lambda_i \, du}} \lambda_i = P_{ij} \lambda_i = q_{ij}.
\end{equation}
Going in the other direction using \eqref{eq:pijIItoI} and remembering that since $Q$ is a generator matrix, $q_{ii} = - \sum_{k \neq i} q_{ij} < 0$, we have,
\[
p_{ij} = \int_0^\infty q_{ij} e^{-\int_0^t \sum_{k \neq i} q_{ik} \, du} \, dt = q_{ij} \int_0^\infty e^{q_{ii} t} \, dt = \frac{q_{ij}}{-q_{ii}}.
\]
Now using \eqref{eq:densTildeTrans} we have,

\[
f_{ij}(t) = -q_{ij} \frac{e^{\int_0^t q_{ii} \, du }}{q_{ij} / q_{ii}} = -q_{ii} e^{q_{ii}t} = \lambda_i e^{-\lambda_i t}.
\]
This is an exponential density and hence it has a constant hazard rate $\alpha_{ij}(t) = \lambda_{i}$.

\noindent
\paragraph{Example 2: Exponential Sojourn Times} 
The example above illustrates that an SMP parameterized via Approach I where $\alpha_{ij}(t)$ is constant in time and also independent of $j$ yields constant (Approach II) transition intensity functions, $\tilde{\alpha}_{ij}(t)$. That is, constant (Approach~I) hazard rates yield constant transition intensity functions, but under the condition that ${\alpha}_{ij}(t)$ is the same for all target states $j$. 

We now show that this condition is necessary for having constant transition intensity functions. To see this, use \eqref{eq:ItoII} with $\alpha_{ij}(t) = \lambda_{ij}$ where for some $j_1$ and $j_2$, $\lambda_{ij_1} \neq \lambda_{i j_2}$. Now,
\begin{equation}
\label{eq:nonConstantTildeAlphaSurrr}
\tilde{\alpha}_{ij}(t)=  p_{ij} \frac{ e^{-\lambda_{ij} t}}{\sum_{k \neq i} p_{ik}  e^{- \lambda_{ik}t}}  \lambda_{ij}.
\end{equation}
Since $\lambda_{ij_1} \neq \lambda_{i j_2}$ we are not able to cancel out the exponent as was done in \eqref{eq:benoitGivingCourseInDenemark}.

Interestingly, the converse does not hold. The previous example showed that time independent transition intensity functions always yield a CTMC and hence constant (Approach I) hazard rates.

\noindent
\paragraph{Example 3: Weibull Distributions}

The Weibull continuous univariate distribution of a non-negative random variable is parameterized by a shape parameter $\eta>0$ and a scale parameter $\mu>0$. It has density, survival function, and hazard rate functions respectively given via,
\[
f(t) = \frac{\eta}{\mu} \Big(\frac{t}{\mu} \Big)^{\eta-1} e^{-(t/\mu)^\eta},
\qquad
S(t) = e^{-(t/\mu)^\eta},
\qquad
\alpha(t) = \frac{\eta}{\mu} \Big(\frac{t}{\mu} \Big)^{\eta-1} .
\]
It is appealing in survival analysis and reliability analysis because if $\eta<1$ the hazard rate is monotonically decreasing; if $\eta=1$ the hazard rate is constant (an exponential distribution) and if $\eta>1$ the hazard rate is monotonically increasing. 

Say now that we are using Approach I and parameterize all of the sojourn time distributions as Weibull distributions, where for transition $i \to j$ we have respective scale and shape parameters $\mu_{ij}$ and $\eta_{ij}$. Now, if we were to consider the Approach II representation, then from \eqref{eq:ItoII} the transition intensity functions are
\begin{equation}
\label{eq:benoitGivingCourseInDenemark1}
\tilde{\alpha}_{ij}(t) = p_{ij} \frac{ e^{-(t/\mu_{ij})^{\eta_{ij}}}}{\sum_{k \neq i} p_{ik}  e^{-(t/\mu_{ik})^{\eta_{ik}}}} \frac{{\eta_{ij}}}{\mu_{ij}} \Big(\frac{t}{\mu_{ij}} \Big)^{{\eta_{ij}}-1} 
= K_{ij}(t) t^{\eta_{ij} - 1},
\end{equation}
where we use the notation $K_{ij}(t)$ to represent all of the components of $\tilde{\alpha}_{ij}(t)$ excluding $t^{\eta_{ij} - 1}$. Now the form of  $K_{ij}(t)$ can determine if the Approach II representation is of a Weibull type or not. A Weibull type representation will follow if $K_{ij}(t)$ is independent of $t$, otherwise it is not.

We now see that if for a fixed $i$ and every $j_1$ and $j_2$ with $j_1 \neq j_2$ we have that 
\begin{equation}
\label{eq:conditionForIandIIWeibullTogether}
\eta_{ij_1} = \eta_{ij_2},
\qquad
\mbox{and}
\qquad
\mu_{ij_1} = \mu_{ij_2},
\end{equation}
then, $K_{ij}(t)$ is independent of $t$ and we obtain a Weibull type Approach II representation with,
\begin{equation}
\label{eq:dudesInBordouxDoWeibull}
\tilde{\alpha}_{ij}(t) = p_{ij} \frac{{\eta_{ij}}}{\mu_{ij}} \Big(\frac{t}{\mu_{ij}} \Big)^{{\eta_{ij}}-1} =  \frac{\eta_{ij}}{p_{ij}^{-1/\eta_{ij}}\mu_{ij}} \Bigg(\frac{t}{p_{ij}^{-1/\eta_{ij}}\mu_{ij}} \Bigg)^{\eta_{ij}-1},
\end{equation}
and hence the scale parameter is modified to $p_{ij}^{-1/\eta_{ij}}\mu_{ij}$ and the shape parameter keeps the same form. Otherwise (if there exists $j_1$ and $j_2$ such that \eqref{eq:conditionForIandIIWeibullTogether} does not hold), then $\tilde{\alpha}_{ij}(t)$ cannot be of the Weibull type as in \eqref{eq:dudesInBordouxDoWeibull}. 

Going the other way, assume we are using Approach II with Weibull type transition intensity functions,
\[
\tilde{\alpha}_{ij}(t) = \frac{\eta_{ij}}{\mu_{ij}} \Big(\frac{t}{\mu_{ij}} \Big)^{\eta_{ij}-1} .
\]
Then by integrating \eqref{eq:pijIItoI} we can in principle obtain $p_{ij}$. It turns out that if for a given state $i$, the shape parameters of all the transition intensity functions is the same (denote it via $\eta_i$) then the integration yields an explicit form,
\begin{equation}
\label{eq:manMan7}
p_{ij} = \frac{\mu_{ij}^{\eta_{i} }}{\sum_{k \neq i} \mu_{ik}^{\eta_{i}} },
\end{equation}
otherwise, there is not an explicit solution. In such a case where $\eta_{ij} = \eta_i$ for all $j$, we also have from equation \eqref{eq:densTildeTrans},
\begin{equation}
\label{eq:fijStopMan}
f_{ij}(t)  =  \frac{\eta_{i}}{\mu_{ij}} \Big(\frac{t}{\mu_{ij}} \Big)^{\eta_{i}-1} \, \frac{ 
e^{
-\sum_{k \neq i} \big(\frac{t}{\mu_{ik}}\big)^{\eta_i}   
}
}
{ \frac{\mu_{ij}^{\eta_{i} }}{\sum_{k \neq i} \mu_{ik}^{\eta_{i}} }}.
\end{equation}
Further, if $\mu_{ij} = \mu_i$ for all $j$ then $p_{ij}$ of \eqref{eq:manMan7} reduces to $\ell_i^{-1}$ where $\ell_i < \ell$ is the number of possible target states from state $i$. Note that this form is not specific to the Weibull type intensity transition function, but will hold whenever the intensity transition functions out of state $i$ are the same.

However, with the Weibull case we can continue further and \eqref{eq:fijStopMan} can be represented as,
\[
f_{ij}(t) =\ell_i \frac{\eta_{i}}{\mu_{i}} \Big(\frac{t}{\mu_{i}} \Big)^{\eta_{i}-1} \, { 
e^{
-\ell_i \big(\frac{t}{\mu_{i}}\big)^{\eta_i}}}
=
 \frac{\eta_{i}}{\ell_i^{-1/\eta_{i}} \mu_{i}} \Big(\frac{t}{\ell_i^{-1/\eta_{i}}\mu_{i}} \Big)^{\eta_{i}-1} \, { 
e^{
- \big(\frac{t}{\ell_i^{-1/\eta_{i}}\mu_{i}}\big)^{\eta_i}}},
\]
which denotes a Weibull distribution with same shape parameter as the transition intensity function and a scale parameter equal to $\ell_i^{-1/\eta_{i}}\mu_{i}$.

\section{Likelihood and Inference}
\label{section:inference}
 
Here we focus on inference for the fully-parametric semi-Markov model defined by the two above mentioned approaches. Survival analysis usually applies for cohort or clinical studies, hence the data is usually gathered from the same subjects repeatedly during a time interval $[0,{\cal T}]$. A key issue in survival analysis and event history analysis is the occurrence of incomplete or sparse observations. For instance, in the case of chronic diseases, when the event of study is death, time of occurrence of this event is not observed for the subjects still alive at time~${\cal T}$. This type of incomplete observation is called \textit{right-censoring}. There are other kinds of incomplete data like left-censoring and interval censoring, see \cite{andersen2005censored,commenges2015dynamical}. \cite{andersen2002multi} present different incomplete data forms which can be handled in multi-state frameworks. Our focus in this paper is right censored data corresponding to when at the end of the observation period, not all individuals under study have reached an absorbing state. 

As our focus is on the fully-parametric case we now present the likelihood functions for both Approach~I and Approach~II. We assume there are $n$ subjects denoted via $h=1,\ldots,n$, for which data is collected and we assume independence between subjects. For each subject, we distinguish between two cases depending on if the subject is in an absorbing state at time ${\cal T}$ or not. {Note that in principle we can set ${\cal T}$ to be subject specific, but for simplicity we do not do so here.} This is recorded via $\delta_h$ where $\delta^{(h)} = 1$ implies no right censoring (subject is in an absorbing state at time ${\cal T}$), and $\delta^{(h)} = 0$ implies right censoring. Further, we record the state evolution denoted via the sequence $\{J\}$ and the sojourn times denoted via the sequence $\{\tau\}$. For simplicity, we assume that all subjects start at the same fixed and known state, denoted via $J_0$. This can be easily generalized.

For subject $h$ during $[0,{\cal T}]$, we use $N^{(h)}$ to denote the number of state transitions up to time ${\cal T}$. The data of the subject is represented via
\[
{\cal H}^{(h)}=\big(J_{0},J^{(h)}_{1}, \ldots, J^{(h)}_{N^{(h)}},  \tau^{(h)}_{1}, \ldots, \tau^{(h)}_{N^{(h)}}  , \delta^{(h)}\big).
\]

Note that if $\delta^{(h)} = 0$ and there is right censoring then we are also interested in the time duration $U^{(h)}$ after the last state is visited. This quantity is represented via,
\begin{equation}
\label{eq:uDefThing}
U^{(h)} = {\cal T} - \sum_{i=1}^{N^{(h)}} \tau^{(h)}_{i}.
\end{equation}

As our subjects are independent we can consider the likelihood contribution of each subject $h$ in isolation, and after denoting it via ${\cal L}^{(h)}$, the full likelihood is
\begin{equation}
\label{eq:oneHundredPercentCorrect}
{\cal L}=\prod_{h=1}^n {\cal L}^{(h)}.
\end{equation}
The specific form of ${\cal L}^{(h)}$ now depends on if we are using Approach~I or Approach~II and as we show below, in Approach~II, it can further be decomposed as in \eqref{eq:decomposeAppII}.

\subsection{Likelihood for Approach~I} 

The form of ${\cal L}^{(h)}$ based on ${\cal H}^{(h)}$ is,
\begin{equation}
\label{eq:approachIlikelihood}
{\cal L}^{(h)}=
\Big(
\prod_{k=1}^{N^{(h)}} p_{J_{k-1} \,J_{k}} \, f_{J_{k-1} \,J_{k}}(\tau_{k}) \Big)\,\Big(S_{J_{N^{(h)}}}(U^{(h)})\Big)^{1- \delta^{(h)}},
\end{equation}
where for brevity we omit the $(h)$ superscripts for the state and sojourn time sequences $\{J\}$ and $\{\tau\}$. To further understand \eqref{eq:approachIlikelihood} consider a recursive construction where each transition without censoring based on $J_{k-1} \to J_k$ with a sojourn time of $\tau_k$ has a likelihood contribution $p_{J_{k-1} \,J_{k}} \, f_{J_{k-1} \,J_{k}}(\tau_{k})$. Further, in case of censoring the last censored transition has likelihood contribution $S_{J_{N^{(h)}}}(U^{(h)})$ based on the survival function of the holding time \eqref{eq:S-i}, corresponding to the last observed state $J_{N^{(h)}}$.

\subsection{Likelihood for Approach II}

We refer to the key relationship \eqref{eq:relation-alpha-alphatilde} and replace $p_{ij}f_{ij}(u)$ in \eqref{eq:approachIlikelihood} with $\tilde{\alpha}_{ij}(u) S_{i}(u)$ to obtain,
\begin{equation}
\label{eq:approach2likelihoodA}
{\cal L}^{(h)}=
\Big(
\prod_{k=1}^{N} \tilde{\alpha}_{J_{k-1} J_k}(\tau_k) S_{J_{k-1}}(\tau_k) 
\Big)
\,\Big(S_{J_{N}}(U)\Big)^{1- \delta},
\end{equation}
where for brevity we omit all superscripts $(h)$. Further, we manipulate the likelihood contribution of each subject as follows:
\begin{equation}
\label{eq:approach2likelihoodC}
\begin{array}{ll}
\displaystyle{\cal L}^{(h)} =
\Big(
\prod_{k=1}^{N} \tilde{\alpha}_{J_{k-1} J_k}(\tau_k) e^{-\int_0^{\tau_k} \tilde{\alpha}_{J_{k-1}}(u) \, du} 
\Big)
\Big(e^{-\int_0^{U} \tilde{\alpha}_{J_N}(u) \, du}\Big)^{1-\delta}
\\[12pt]
\,\qquad=
\displaystyle \prod_{k=1}^{N+1} \big(\tilde{\alpha}_{J_{k-1} J_k}(\tau_k)\big)^{ \mathbbm{1}_{\{k\neq N+1\}}} \, e^{-\int_0^{\tau_k} \tilde{\alpha}_{J_{k-1}}(u) \, du}. 
\end{array}
\end{equation}

In the first line we use the representation in \eqref{eq:SiFromAlphaTilde}. For the second line, by increasing the summation from $N$ to $N+1$ we are able to remove $\delta$ by defining $\tau_{N+1} = U$, $J_{N+1} \equiv -1$, and using the fact that $\tilde{\alpha}_{ij}(u) \equiv 0$ in cases where $i$ is an absorbing state and $0^0 \equiv 1$.

Now, we are able to separate ${\cal L}^{(h)}$ into the form
\begin{equation}
\label{eq:decomposeAppII}
{\cal L}^{(h)} = \prod_{i=1}^\ell \prod_{j=1}^\ell {\cal L}^{(h)}_{ij},
\end{equation}
following similar ideas as those initially presented in \cite{hougaard1999multi}. To achieve such a separation, it is useful to define the indicators
\[
\delta^{k-1}_{i \to j} =\mathbbm{1}_{\{J_{k-1}=i,\, J_k=j\} },
\qquad 
\mbox{and} 
\qquad
\delta^{k-1}_{i \not \to j}  = \mathbbm{1}_{\{J_{k-1}=i,\, J_k \neq j\}}.
\]
We may now manipulate \eqref{eq:approach2likelihoodC} to obtain
\begin{equation}
\label{eq:niceNiceCorrectEquation}
{\cal L}^{(h)}_{ij} =
 \prod_{k=1}^{N+1}\Big(\tilde{\alpha}_{ij}(\tau_k) e^{-\int_0^{\tau_k}  \tilde{\alpha}_{ij}(u)\,du}\Big)^{\delta^{k-1}_{i \to j} } 
\Big(
e^{-\int_0^{\tau_k}  \tilde{\alpha}_{ij}(u)\,du}\Big)^{\delta^{k-1}_{i \not \to j}}.
\end{equation}

The form of \eqref{eq:oneHundredPercentCorrect}, \eqref{eq:decomposeAppII}, and \eqref{eq:niceNiceCorrectEquation} allows us to treat each transition separately as if each transition intensity transition function $\tilde{\alpha}_{ij}(\cdot)$ has its own set of parameters. 

\subsection{Parametric forms and Covariate Information}
 \label{sec:thePlaceToBe}
 
In carrying out inference, we assume a parametric form for $\alpha_{ij}(\cdot)$ in Approach~I or $\tilde{\alpha}_{ij}(\cdot)$ in Approach~II. We also allow for covariate information where the natural way to introduce covariates in a multi-state model is a \textit{Cox-like proportional hazard model} which can be defined by using either the hazard function of the sojourn times or by intensity transitions, see   \cite{meira2009multi}. Hence for a vector of covariates $Z$, we have
\begin{equation}
\label{Eq:covariate-sojourn}
 {\alpha_{ij}}(t ~|~ Z)={\alpha}^{({\theta}_{ij})}_{ij,0}(t) e^{\beta_{ij}^TZ},
 \end{equation}
 for Approach~I. Further we have
 \begin{equation}
 \label{Eq:covariate-approachII}
  \tilde{\alpha}_{ij}(t~|~ Z)={\tilde{\alpha}}^{(\tilde{\theta}_{ij})}_{ij,0}(t) e^{\tilde{\beta}_{ij}^TZ},
\end{equation}
for Approach~II.

Here ${\alpha}^{({\theta}_{ij})}_{ij,0}(\cdot)$ and  ${\tilde{\alpha}}^{(\tilde{\theta}_{ij})}_{ij,0}(\cdot)$ are the baseline functions for the hazard rate and transition intensity functions respectively. They each follow a parametric form, determined via $\theta_{ij}$ and $\tilde{\theta}_{ij}$ respectively. Further, $\beta_{ij}$ and $\tilde{\beta}_{ij}$ are the regression parameters for transition $i \to j$ associated with the covariates of the subject. Observe that there are major differences in the interpretation of the regression coefficients $\beta_{ij}$ of Approach~I and $\tilde{\beta}_{ij}$ of Approach~II. The { former determine} the hazard ratios dealing with sojourn times and bear no effect on the actual transitions of the SMP. The latter, determine the hazard ratios dealing with risks and also affect the instantaneous rate of transitioning. 
 
When estimating the parameters for such a model, with Approach~I, we also need to estimate $p_{ij}$, whereas with Approach~II this is not needed as $p_{ij}$ is implicitly determined via \eqref{eq:pijIItoI}.  Interestingly, via Approach~II, we have that the covariates implicitly affect $p_{ij}$ ,whereas with Approach~I, we may wish to set $p_{ij}(Z)$ using multinomial regression, however it isn't common practice.

\subsection{Contrasting Inference for Approach~I and Approach~II}

 The key difference in the likelihood expressions between Approach~I \eqref{eq:approachIlikelihood} and Approach~II \eqref{eq:decomposeAppII} and \eqref{eq:niceNiceCorrectEquation} is in computational tractability. As noted by \cite{Krol} the optimization of the likelihood for Approach~I could be challenging when the number of covariates and parameters is large due to a complex form of the sojourn time distribution. This is especially the case when the number of variables is large compared to the size of the data. It should further be noted that modeling using Approach~I requires the embedded chain parameters in addition to the parameters involved in \eqref{Eq:covariate-sojourn}. This means that for a model with $\ell$ states, there may be up to $\ell^2 -\ell$ more parameters to estimate when using Approach~I, in comparison to Approach~II, where all the parameters appear in \eqref{Eq:covariate-approachII}.
 
With Approach~II, the decoupling in \eqref{eq:decomposeAppII} allows to optimize the likelihood for each transition $i \to j$ separately if each hazard intensity transition function \eqref{Eq:covariate-approachII} has its own set of parameters $\tilde{\theta}_{ij}$ and $\tilde{\beta}_{ij}$. In such a framework, the full likelihood of the SMP can be maximized separately by maximizing the likelihood of the different separate two-state models, see \cite{meira2009multi}. Also, the inference of the SMP can be easily handled by using survival models like the Cox model as the separate likelihood arising in survival analysis. 
 
Hence Approach~II enables fitting SMP models using survival analysis methods and software, see for example \cite{therneau2000cox,cook2007statistical,jackson2016flexsurv}. This also applies to the case where extensions to the proportional hazard assumption are needed. The forms in \eqref{Eq:covariate-sojourn} and \eqref{Eq:covariate-approachII} are clearly based on a proportional hazard assumption with fixed covariates where there is only a multiplicative effect on the baseline hazard functions independent of time. An extension is to consider time-dependent covariates $Z(t)$ as in \cite{andersen2000multi,andersen2002multi}. Further, in both proportional models the effect of variables is assumed to have a linear (or log-linear) functional form. A more flexible model is to consider non-linear effects using smooth functions of the covariates such as the Generalized Additive Model (GAM) presented by  \cite{hastie1990exploring}. With such extensions, using Approach~II is much more straightforward than using Approach~I due to the decoupling and because there is ample statistical software available for survival analysis.

One may also wish to incorporate random effects in the models using frailty models. In this case, both \eqref{Eq:covariate-sojourn} and \eqref{Eq:covariate-approachII} could also include random effect to take into account the correlation between observations/subjects. This has been handled in \cite{liquet2012investigating} for Approach~II by exploiting the decoupling of the likelihood. Such random effects may also be incorporated in Approach~I, however the computational effort will be greater.
 
\section{Semi-Markov Application in Practice}\label{sec:application}

In this section, we illustrate the inference  of the  semi-Markov model on two real datasets. {Our purpose with these examples is to help practitioners  see how the mathematical details presented above interact with statistical software. In light of this, we chose datasets that have been analyzed previously with a one sided view (either Approach~I or Approach~II but not both). These datasets are readily available in the R ecosystem.} All the numerical results including tables and figures, are freely reproducible through R codes in a comprehensible detail vignette available in \cite{Liq_github}. For both datasets, we compare results of full parametric models based on Approach~I where we estimate sojourn time distributions and transition probabilities of the embedded chain, and based on Approach~II where we estimate transition intensity functions. 

To carry out estimation using Approach~I we use the { \texttt{SemiMarkov}} package, see \cite{listwon2015semimarkov}. This package implements inference using the likelihood \eqref{eq:oneHundredPercentCorrect} with \eqref{eq:approachIlikelihood} while allowing to use different parametric forms for the sojourn time distribution including exponential, Weibull, and exponentiated Weibull, see \cite{foucher2005semi}. It also supports covariates with standard inference for the covariate coefficients $\beta_{ij}$, including the Wald test.  It allows the possibility to include different covariates for each transition. While the focus of the inference with this package is Approach~I, as output, the package can automatically provide the intensity transition functions $\tilde{\alpha}_{ij}(\cdot)$, referred to as the ``hazard function of the SMP''.

For Approach~II, the decoupling in \eqref{eq:decomposeAppII} with \eqref{eq:niceNiceCorrectEquation} allows us to use any package that implements inference for survival analysis since for each transition $i \to j$ we can maximize,
\[
{\cal L}_{ij} = \prod_{h=1}^n {\cal L}_{ij}^{(h)}
\]
separately if each hazard intensity transition function \eqref{Eq:covariate-approachII} is assumed to have its own set of parameters $\tilde{\theta}_{ij}$ and $\tilde{\beta}_{ij}$. For this we can use different R packages including \texttt{flexsurv} (\cite{jackson2016flexsurv}), \texttt{survival} (\cite{survival-package}), and \texttt{eha} (\cite{ehapack}). In the examples here, we mainly use \texttt{flexsurv} which offers an easy way to run different parametric forms for $\tilde{\alpha}_{ij}(\cdot)$ including exponential, Weibull, gamma, and generalized gamma (see \cite{Prent1975}) among many other forms. It also supports covariates with standard inference for the covariate coefficients $\tilde{\beta}_{ij}$.

We note that the purpose of this section is not to be an extensive survey of all of the R packages that can be used for inference in multi-state models. For a comprehensive software survey, see the survival task view from R (\cite{survivaltaskview}). Our focus here is to illustrate how the aforementioned packages can be used to easily carry out inference, while we also illustrate a few key points.

\subsection{Progressive Three-State Model for the Stanford Heart Transplant data}

As an illustrative example, we revisit the analysis of the Stanford Heart Transplant data, freely available through the \texttt{survival} package. A full description of the data can be found in \cite{Crow77}. The dataset presents the survival of patients on the waiting list for the Stanford heart transplant program. We analyze the data similarly to \cite{meira2011p3state} which use their R package \texttt{p3state} and propose an illness-death model similar to  Figure~\ref{Fig:IllD}. Their states  represent  ``alive without transplant'', ``alive with transplant'', and ``dead''. 

We consider the same dataset, where patients' records from October 1967 (start of the program) until April 1974 includes the information of $103$ patients where $69$ received heart transplants, and $75$ died. Three fixed covariates for each patient are available. These include the age at acceptance (\texttt{age}), the year of acceptance (\texttt{year}), and previous surgery (\texttt{surgery}: coded as \texttt{1} = yes; \texttt{0} = no).

\begin{figure}[h!]
\center
\includegraphics[scale=0.3]{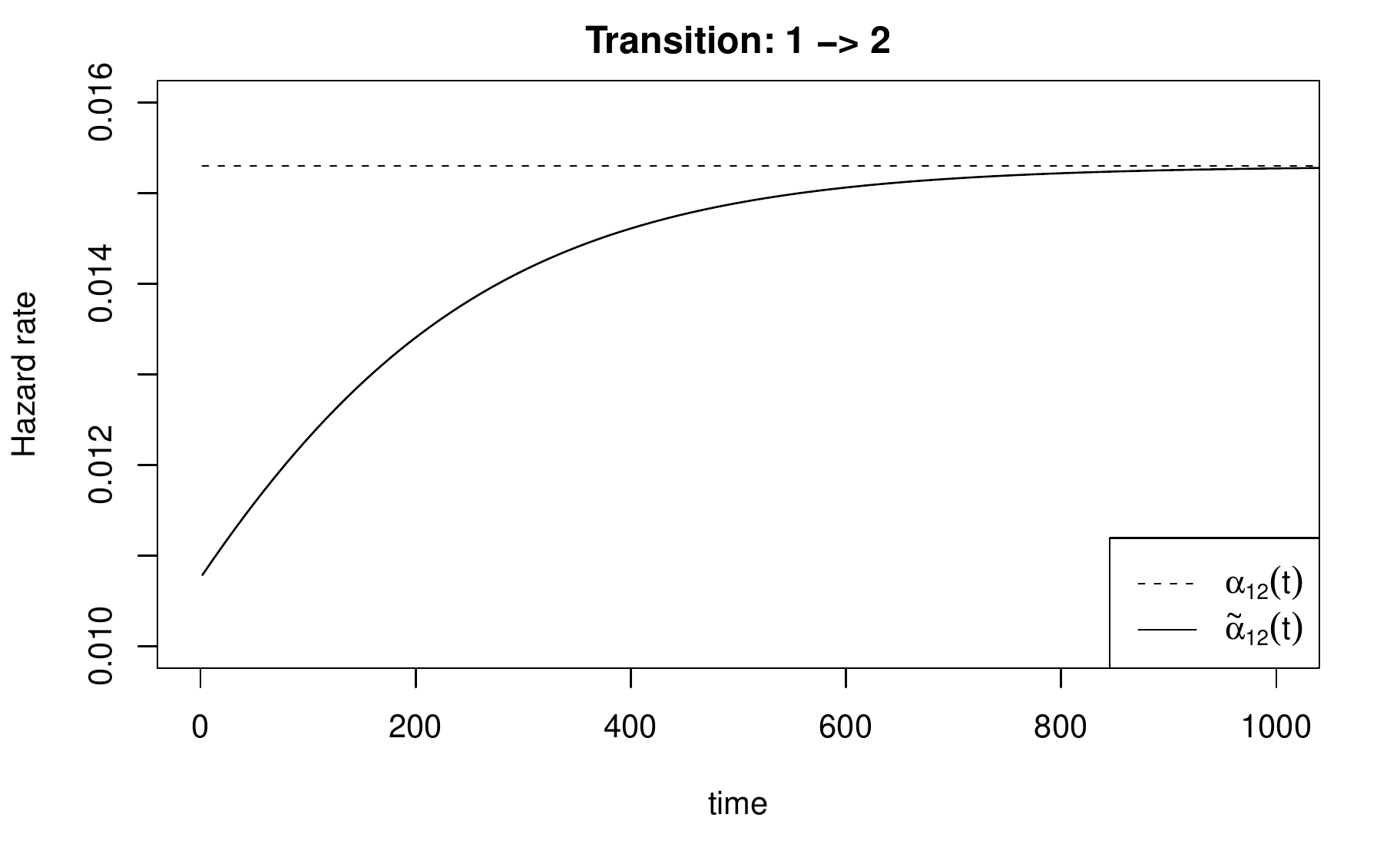}
\includegraphics[scale=0.3]{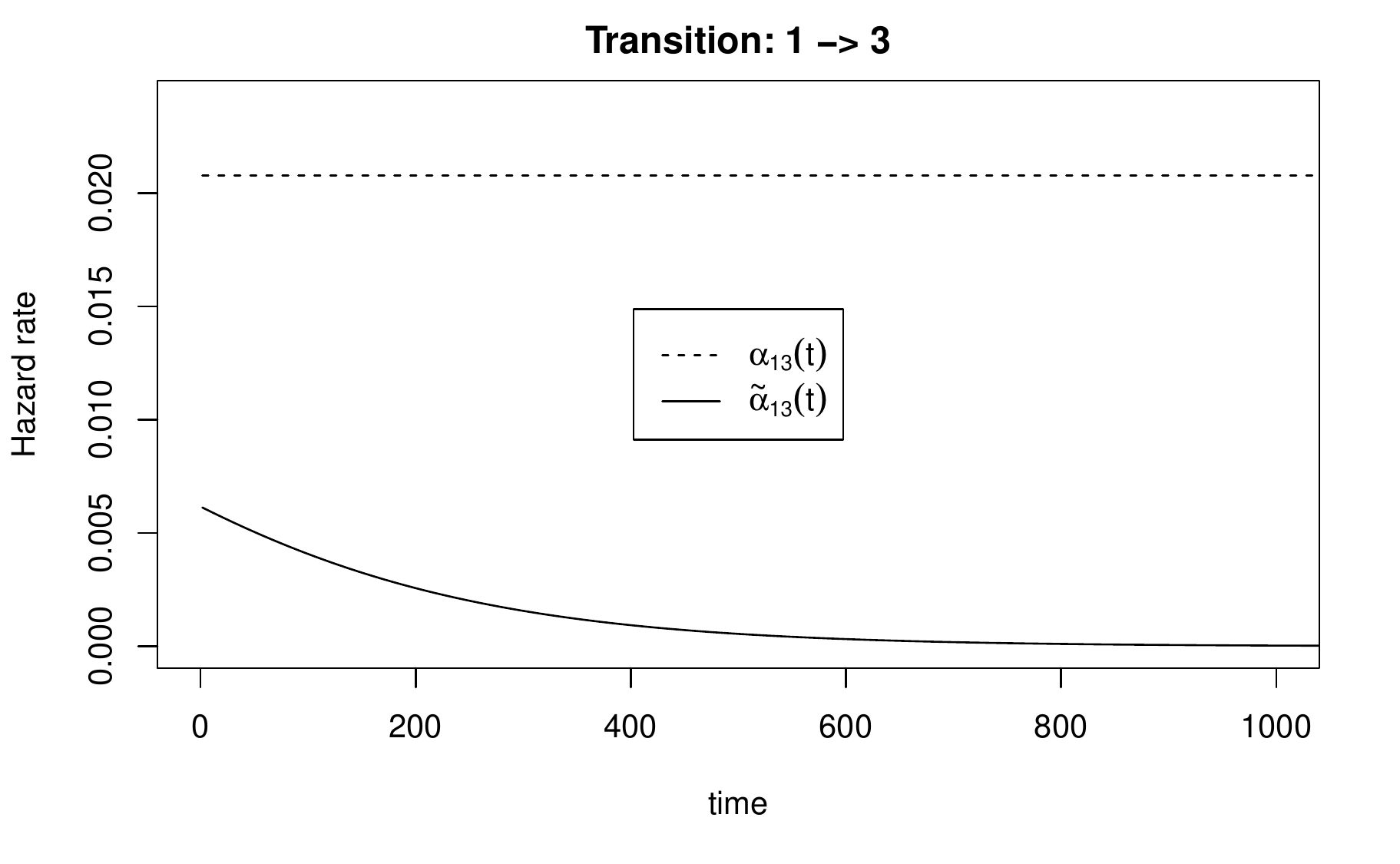}
\caption{
\small Baseline hazard rate of the sojourn times (Approach-I: $\alpha_{ij}(\cdot)$) and baseline hazard rate of the semi-Markov process (Approach-II: $\tilde{\alpha}_{ij}(\cdot)$) for transition $1\to 2$ (top plot) and $1\to 3$ (bottom plot)
}
\label{Fig:hazardAppli1}
\end{figure}

We first use the data without taking into account any covariate effects to show that an exponential distribution on the baseline hazard rate of the sojourn time, $\alpha_{ij}(\cdot)$ does not produce constant intensity transition functions, $\tilde{\alpha}_{ij}(\cdot)$. This point was shown theoretically in Example~2, of Section~\ref{sec:examples} and we now illustrate it using data and the { \texttt{SemiMarkov}}  package. Using the package we obtain both $\alpha_{ij}(\cdot)$ and $\tilde{\alpha}_{ij}(\cdot)$ for $(i,j) = (1,2)$ and $(i,j) = (1,3)$. The results are in Figure \ref{Fig:hazardAppli1} where we see that while $\alpha_{ij}(\cdot)$ are constant, $\tilde{\alpha}_{ij}(\cdot)$ are clearly not constant across time. These curves in fact follow a form like \eqref{eq:nonConstantTildeAlphaSurrr}. 

We now continue to analyze the data fully with Approach~I using the { \texttt{SemiMarkov}} package where we compare different parametric forms and covariates in the model. Table~1
 presents the inference results of $6$ different models investigated depending on the distribution and covariates chosen for each transition. For each model we present the $\hat{\beta}_{ij}$ estimates and their estimated standard error. We also point out that the { \texttt{SemiMarkov}}  package yields $p$-values for the associated Wald tests. We also present the estimates of the transition probabilities of the embedded chain, $\hat{p}_{ij}$. It is evident that these are outputs from likelihood estimation as they don't fully agree across models (they are not proportion estimates). For the final two models, {\em Weibull + Select} and {\em Weibull/Exp + Select} are based on model selection of the significant covariates with more details in the vignette, see \cite{Liq_github}. Observe that the final model, {\em Weibull/Exp + Select}, incorporates two different distributions.

\begin{figure}[h!]
\center
\includegraphics[scale=0.7]{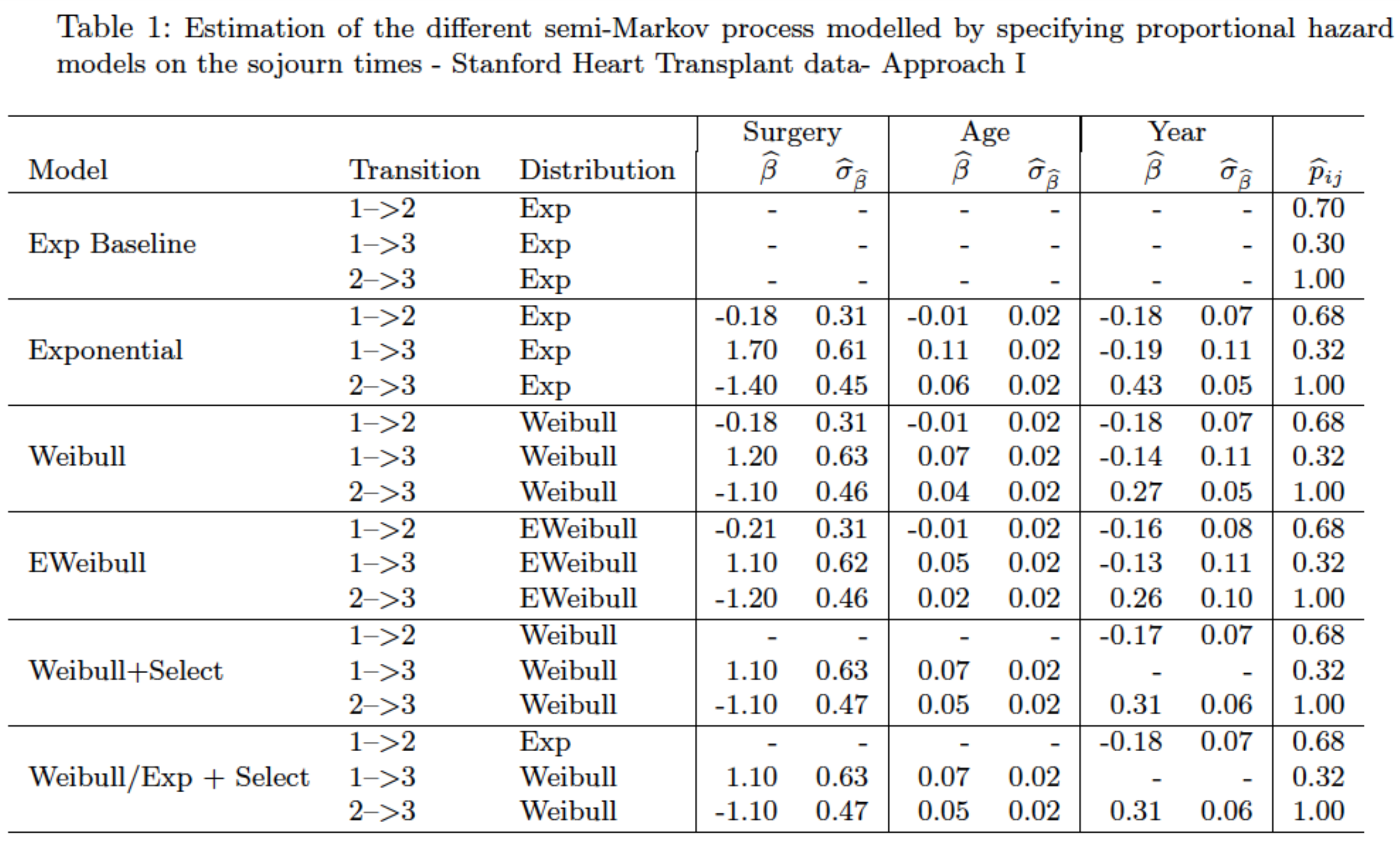}
\end{figure} 


For choosing between the different proposed models, we use the \textit{Expected Kullbak-Leibler} (EKL) risk as in  \cite{liquet2012investigating}. This is done via the \textit{Akaike Information Criterion} (AIC) which is an estimate of the EKL risk in a parametric framework using maximum likelihood method, see \cite{liquet2003bootstrap,commenges2008estimating}. For each of the $6$ models described above, we obtain the AIC using output from { \texttt{SemiMarkov}}  and present it in the first part of Table~2.

The best model according to the AIC is given by the {\em Exponentiated Weibull} model. However, due to  convergence issues during the optimization (see the vignette \cite{Liq_github}), we select the second best result given by the {\em Weibull/Exp + Select} model as the most appropriate model for this dataset according to AIC using Approach~I.

\begin{figure}[h!]
\center
\includegraphics[scale=0.7]{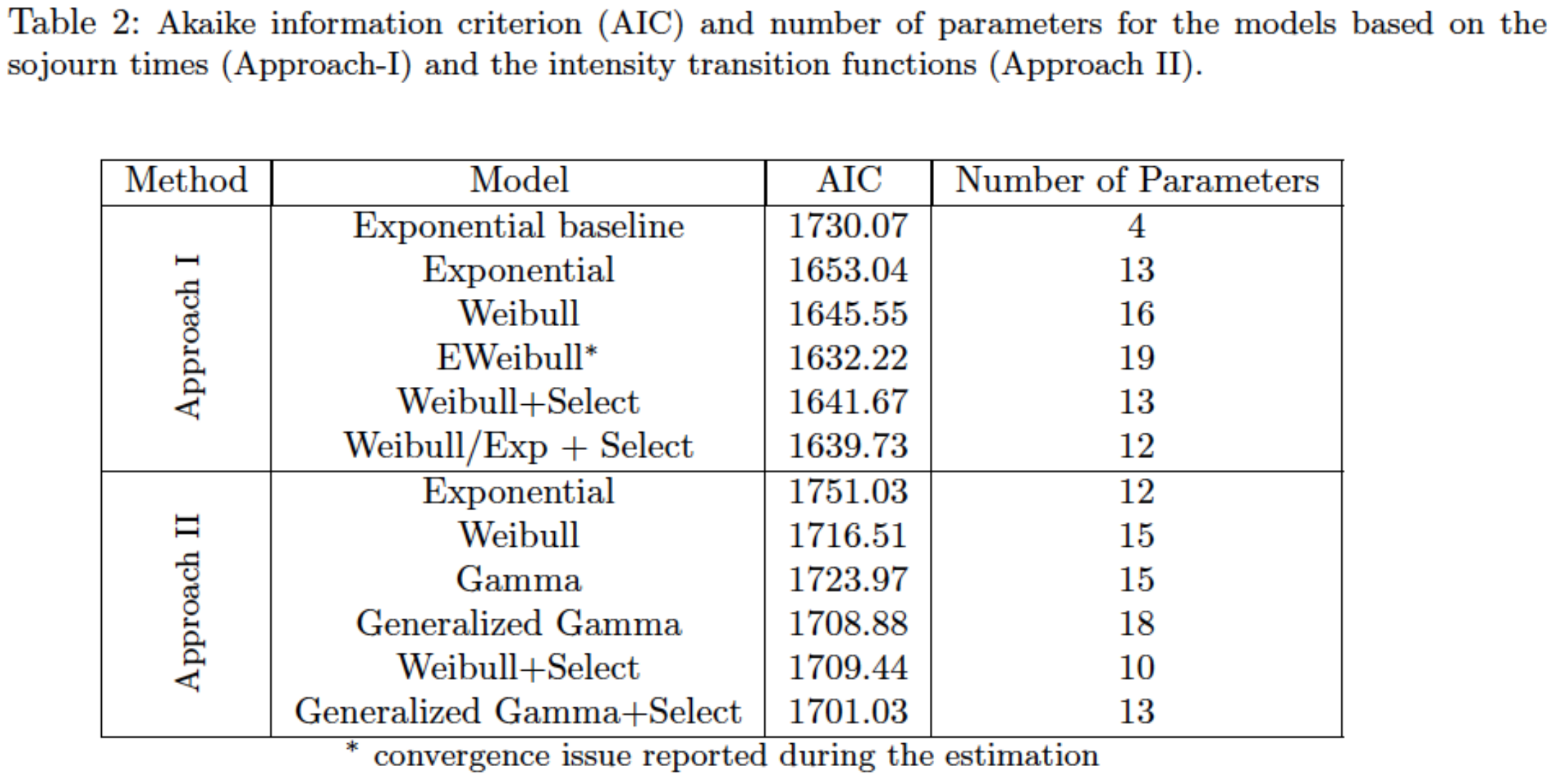}
\end{figure} 
%
%
%


Moving onto inference using Approach~II, we investigate and estimate different models by specifying different proportional intensity transition models as defined in equation \eqref{Eq:covariate-approachII}. We use \texttt{flexsurv} to estimate $\tilde{\alpha}_{ij}(\cdot)$ for several models which we present in the second part of Table~2.
The models {\em Exponential}, {\em Weibull}, {\em Gamma}, and {\em Generalized Gamma} are all estimated with all three covariates in the model. The additional two models, {\em Weibull+Select}, and {\em Generalized Gamma + Select}, have a reduced set of covariates with a procedure consistent with that used for Approach~I above. More details are in the vignette (see \cite{Liq_github}). The AIC based performance of each of these models is presented in the second part of Table~2.
 
The best model according to the AIC is {\em Generalized Gamma + select}. In this model only the \texttt{age} covariate  {is  used} for transition $1 \to 2$, only the \texttt{year} covariate is used for transition $1\to 3$, and both the \texttt{surgery} and \texttt{age} covariates are used for transition $2 \to 3$. 
For this dataset, the best model based on Approach~I presents a better AIC value ($1639.73$) in comparison to the best model based on Approach~II ($1701.03$). 

As discussed in Section~\ref{sec:thePlaceToBe} the regression coefficients of Approach~I and those of Approach~II do not have the same interpretation. Indeed for the hazard rate of the sojourn time, the regression coefficients can be interpreted in terms of relative risk on the waiting time (i.e.\, given an $i \to j$ transition). {For example the positive coefficient for 'surgery' for the $1 \to 3$ transition, as appearing in Table~1, can be interpreted as an indication that death without getting a heart transplant is less favorable for those that had previous surgery}. While the hazard rate of the SMP based on transition intensity functions can be interpreted as the subject's risk of passing from state $i$ to state $j$. {Interpretation of the regression coefficients using both Approach~I and Approach~II is further discussed in the second example below.}

\begin{figure}[h]
\center
\includegraphics[scale=0.25]{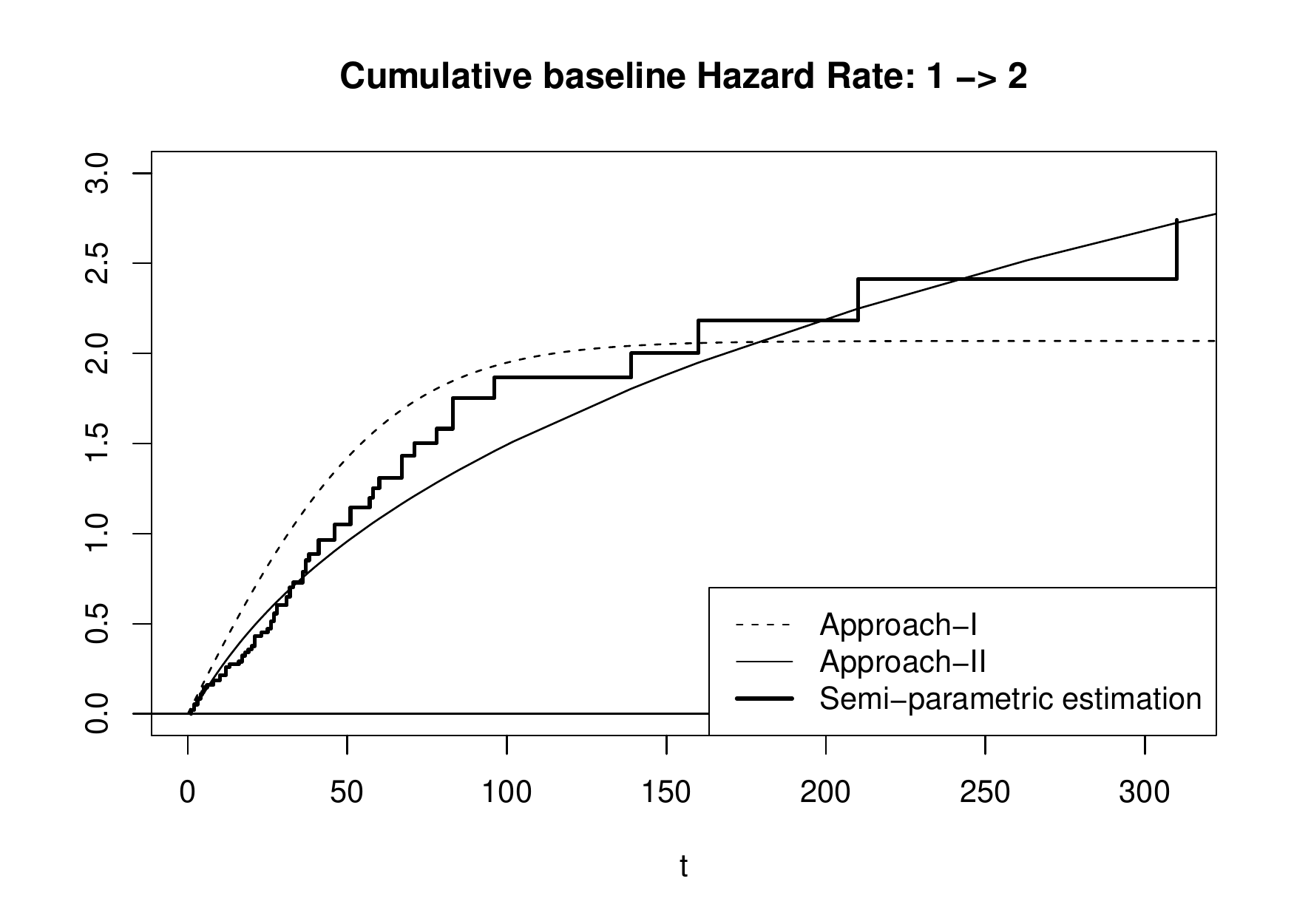}
\includegraphics[scale=0.25]{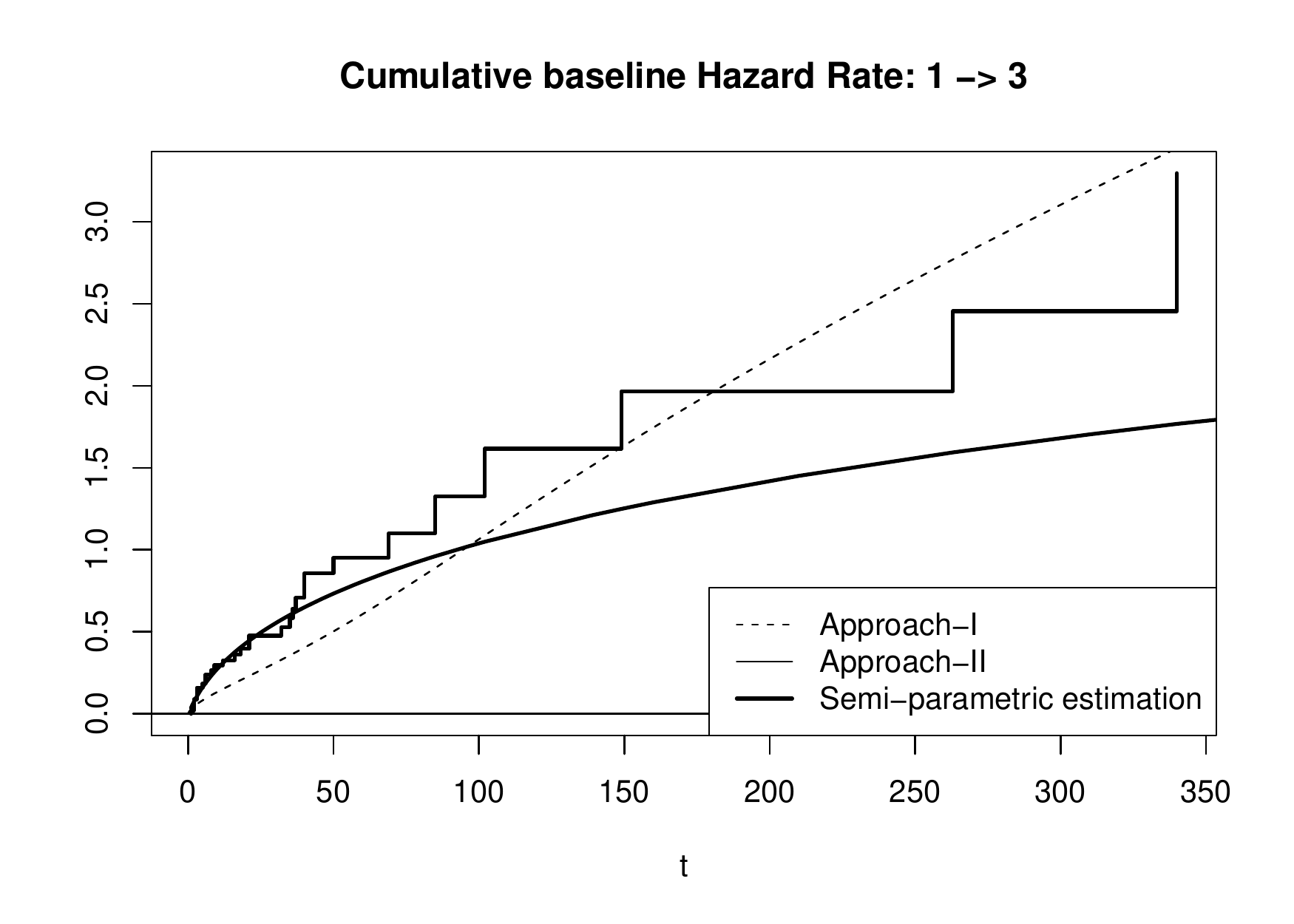}
\includegraphics[scale=0.25]{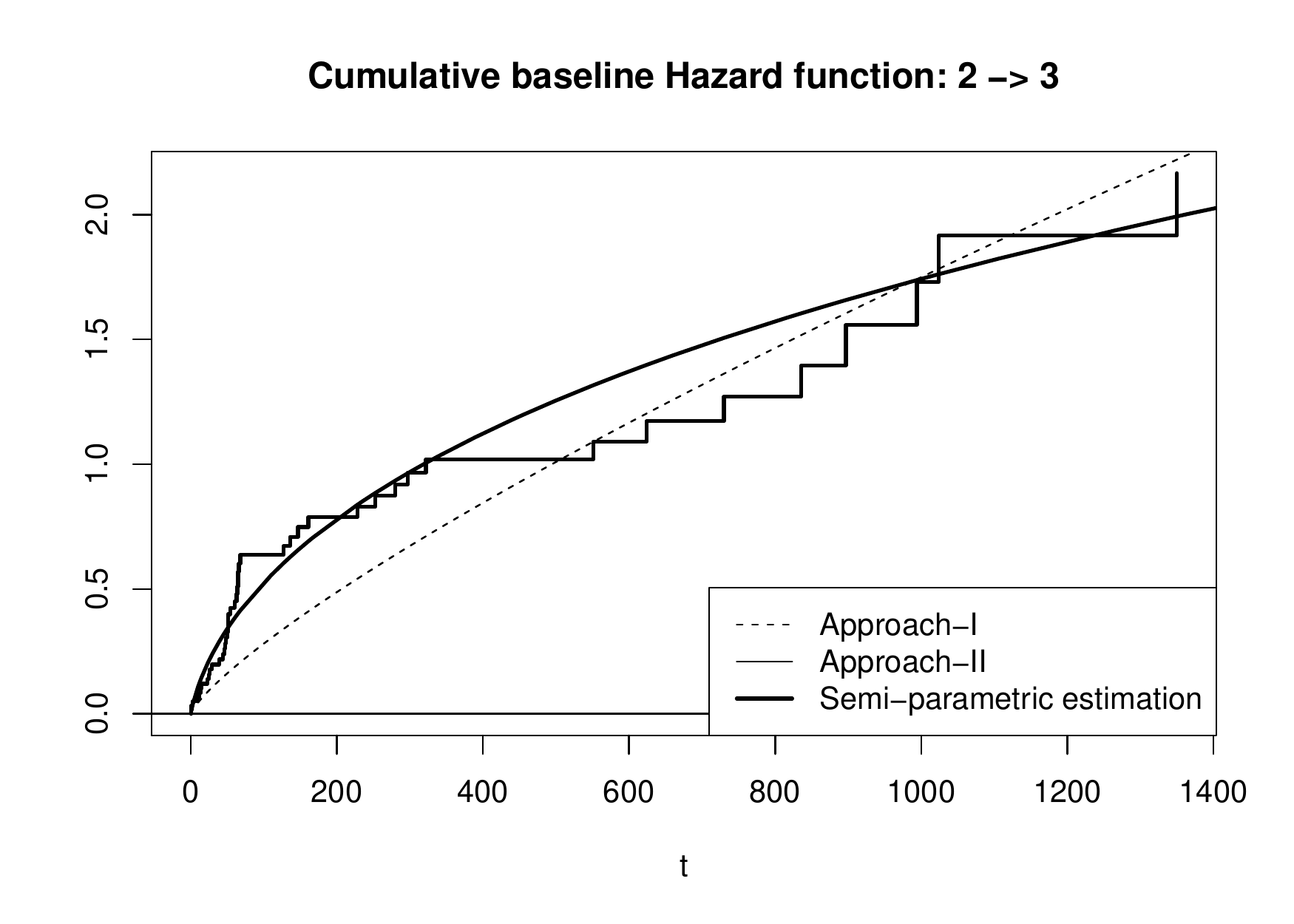}
\caption{\small Cumulative of the baseline hazard rate of the SMP (intensity transition function) estimated from best AIC model of Approach-I and Approach-II.  A semi-parametric estimation is presented as benchmark}
\label{Fig:cumul}
\end{figure}

Since a key object of a Semi-Markov process is the transition intensity function, generally called the hazard rate of the SMP, it is also useful to visualize estimates of these functions based on both Approach~I and Approach~II. Further insight may be gained by plotting these against  non-parametric estimators. For this, we used the  Breslow estimator \cite{breslow1972discussion}, which yields an estimate for the baseline cumulative risk function of a cox-regression type model of each transition. Figure \ref{Fig:cumul} plots the cumulative baseline (all covariates are set to $0$) estimated transition intensity functions, estimated via Approach~I and Approach~II, against the Breslow estimator. These cumulative functions are based on the parameter estimates for the best model of Approach~I ({\em Weibull/Exp + Select}) vs. the best model of Approach~II ({\em Generalized Gamma + Select}). In this case, the plots in Figure \ref{Fig:cumul} hint at a better fit (closer to the non-parametric estimation) of the SMP using Approach~I in accordance with the results of the AIC values discussed above.

\subsection{Reversible Semi-Markov Model for the Asthma Control Data}

\begin{figure}
\center
\includegraphics[scale=0.6]{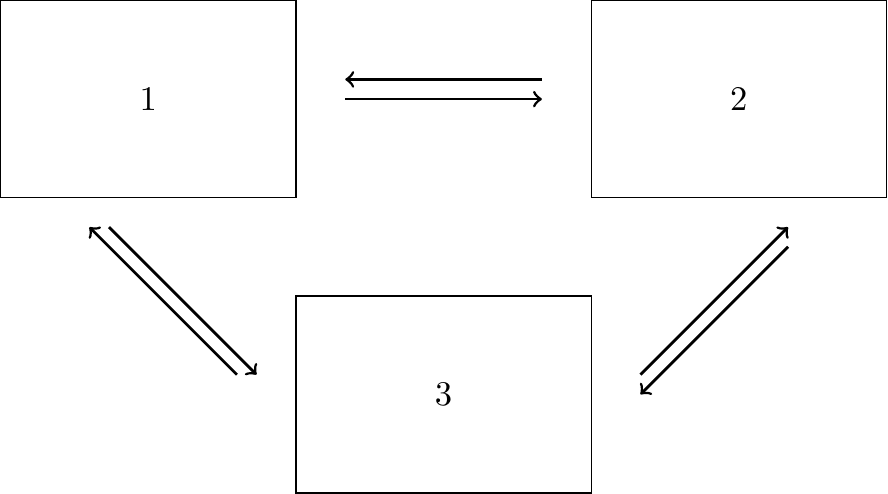}
\caption{{{\small The Reversible three-state Model. }}}\label{Fig:rev3}
\end{figure}

We revisit the analysis of the asthma control data which has been used to illustrate the { \texttt{SemiMarkov}} R package in \cite{listwon2015semimarkov}. {See also the related papers using the same data,\cite{saint2006overweight} and \cite{saint2003analysis}.} The data consists of the follow-up of severe asthmatic patients consulting at varied times according to their perceived
needs. At each visit, several covariates were recorded and asthma was evaluated. The available  data (\texttt{asthma}) from the \texttt{SemiMarkov} package presents a random selection of the $371$ patients with at least two visits. In this illustration, we only use the BMI (Body Mass Index) covariate coded: \texttt{1} if BMI $\geq 25$ and \texttt{0} otherwise. Similarly to the analysis in \cite{listwon2015semimarkov}, we use a reversible three-state model as presented in Figure~\ref{Fig:rev3}. This investigates the evolution of asthma based on, ``optimal control'' (State 1), ``sub-optimal control'' (State 2) and ``unacceptable control'' (State 3). More details about concepts of control scores can be found in \cite{saint2003analysis}. Observe that in this model, none of the states are absorbing and hence the model is said to be {\em reversible}. 

{We note that this data is based on scheduled visits every three months or according to patients' perceived needs, \cite{saint2006overweight},  and is thus potentially interval censored since the exact times of state transitions are approximated to be at visit times. This aspect of the analysis was not the focus of \cite{listwon2015semimarkov} and the related papers. While dealing with such interval censoring is important we do not focus on it here further and refer the reader to \cite{commenges2002inference} and \cite{van2016multi}.}





\begin{figure}[h!]
\center
\includegraphics[scale=0.22]{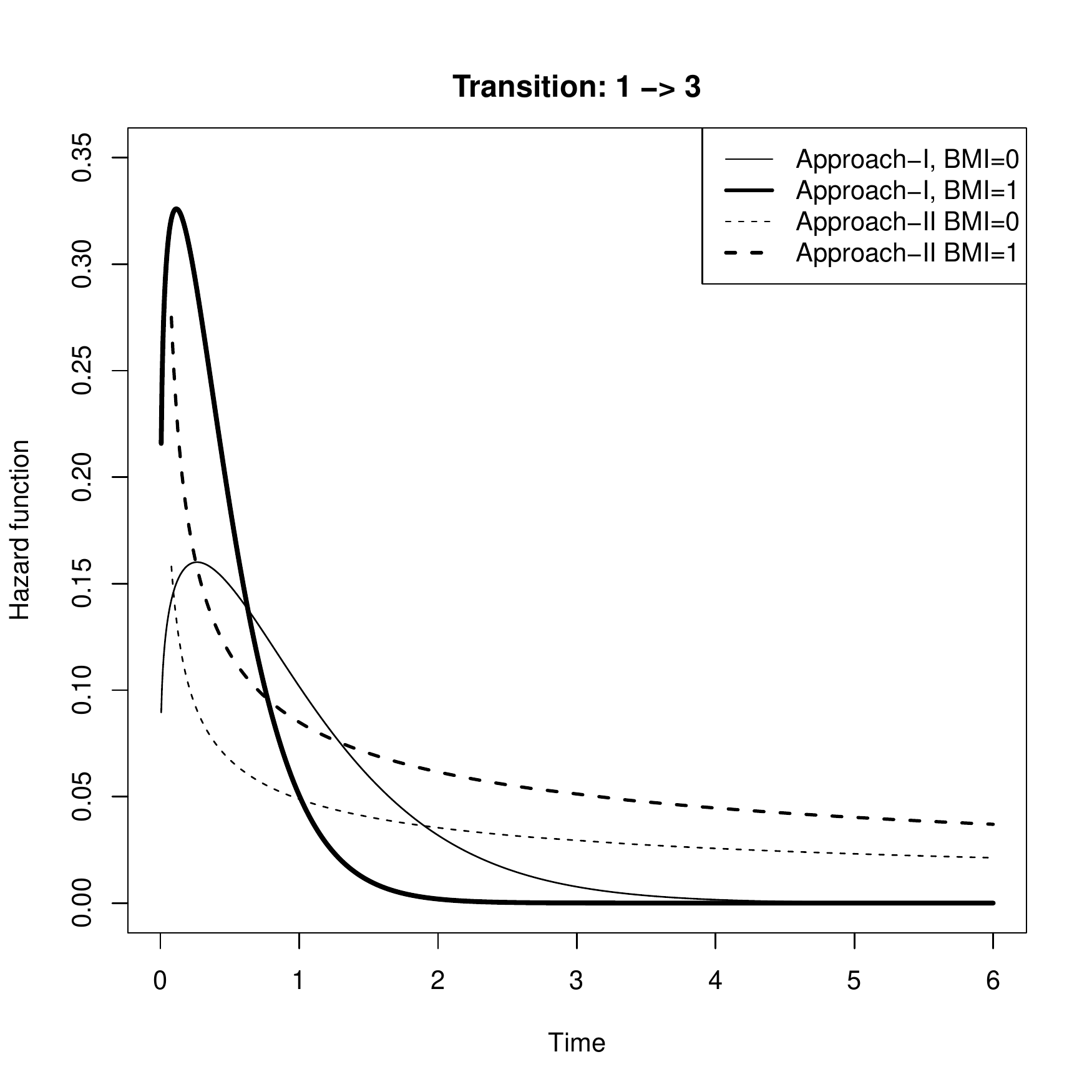}
\includegraphics[scale=0.22]{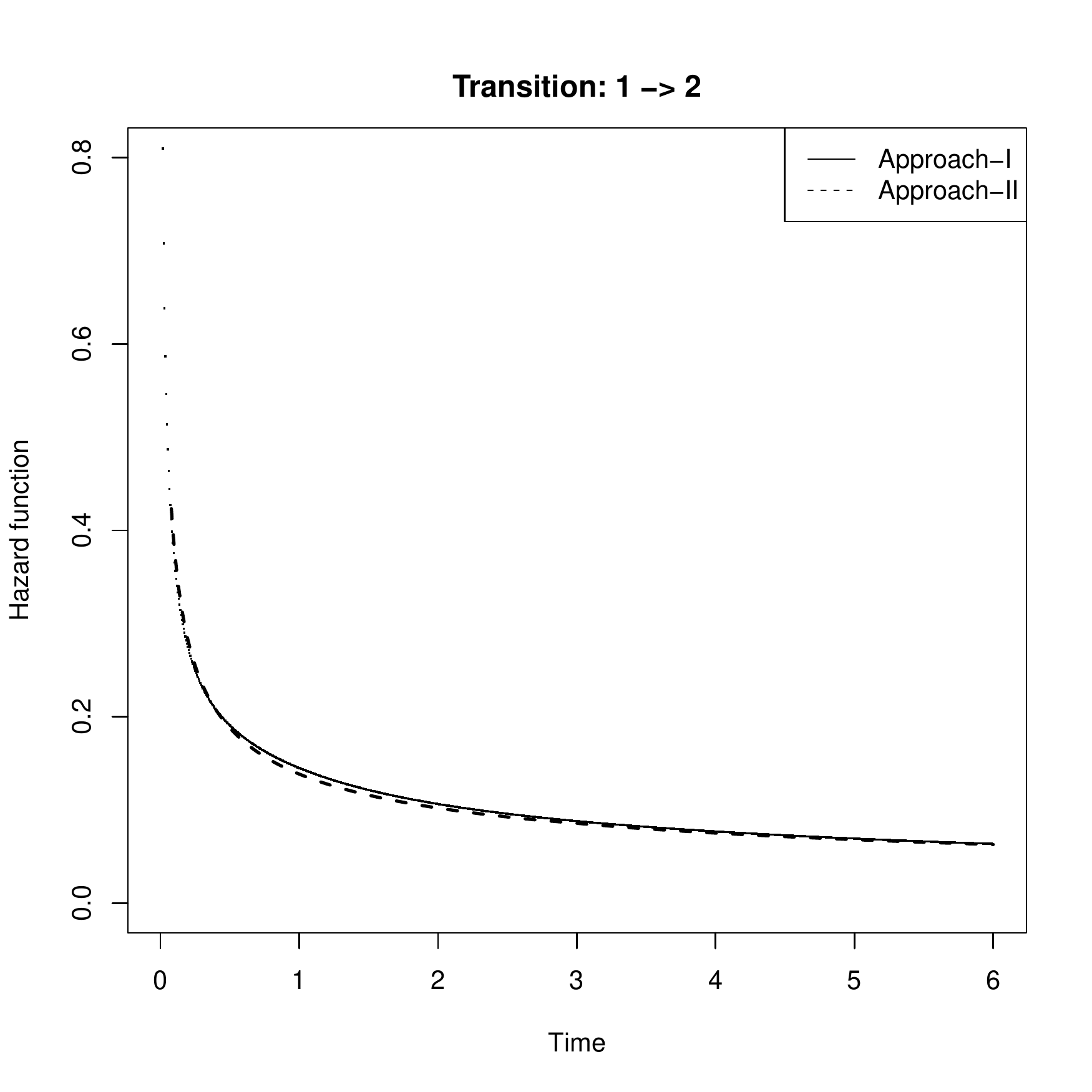}
 \includegraphics[scale=0.22]{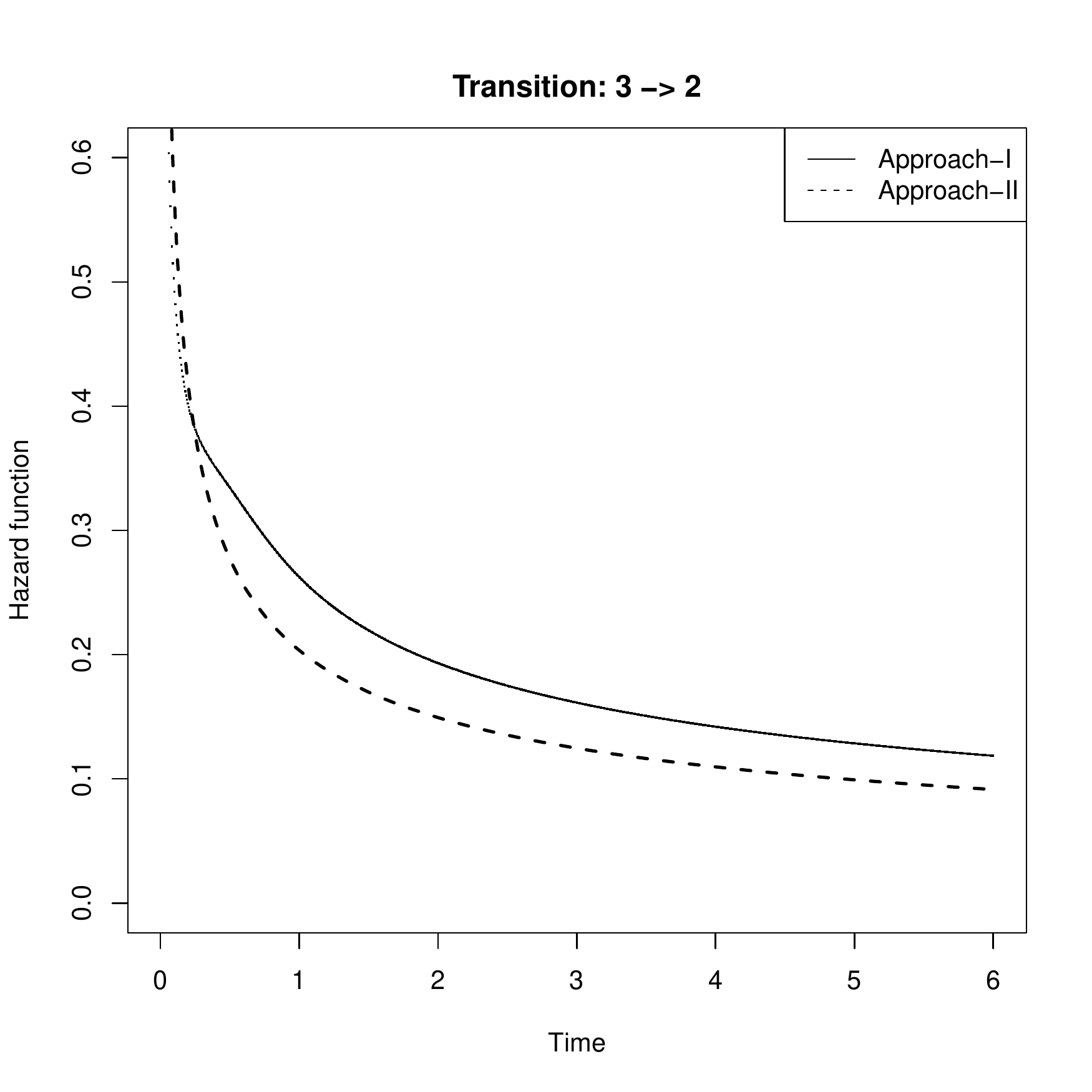}
\includegraphics[scale=0.22]{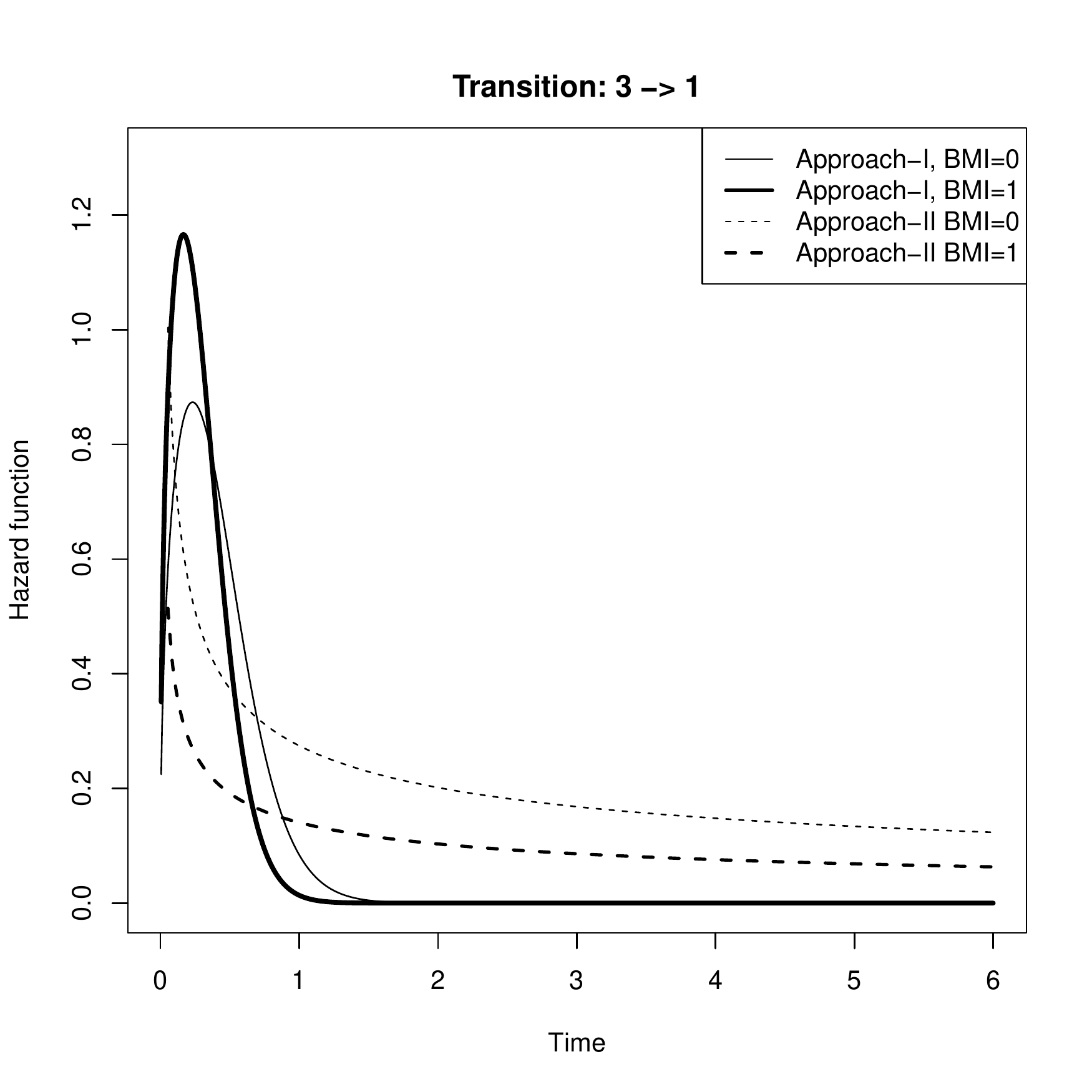}
 \includegraphics[scale=0.22]{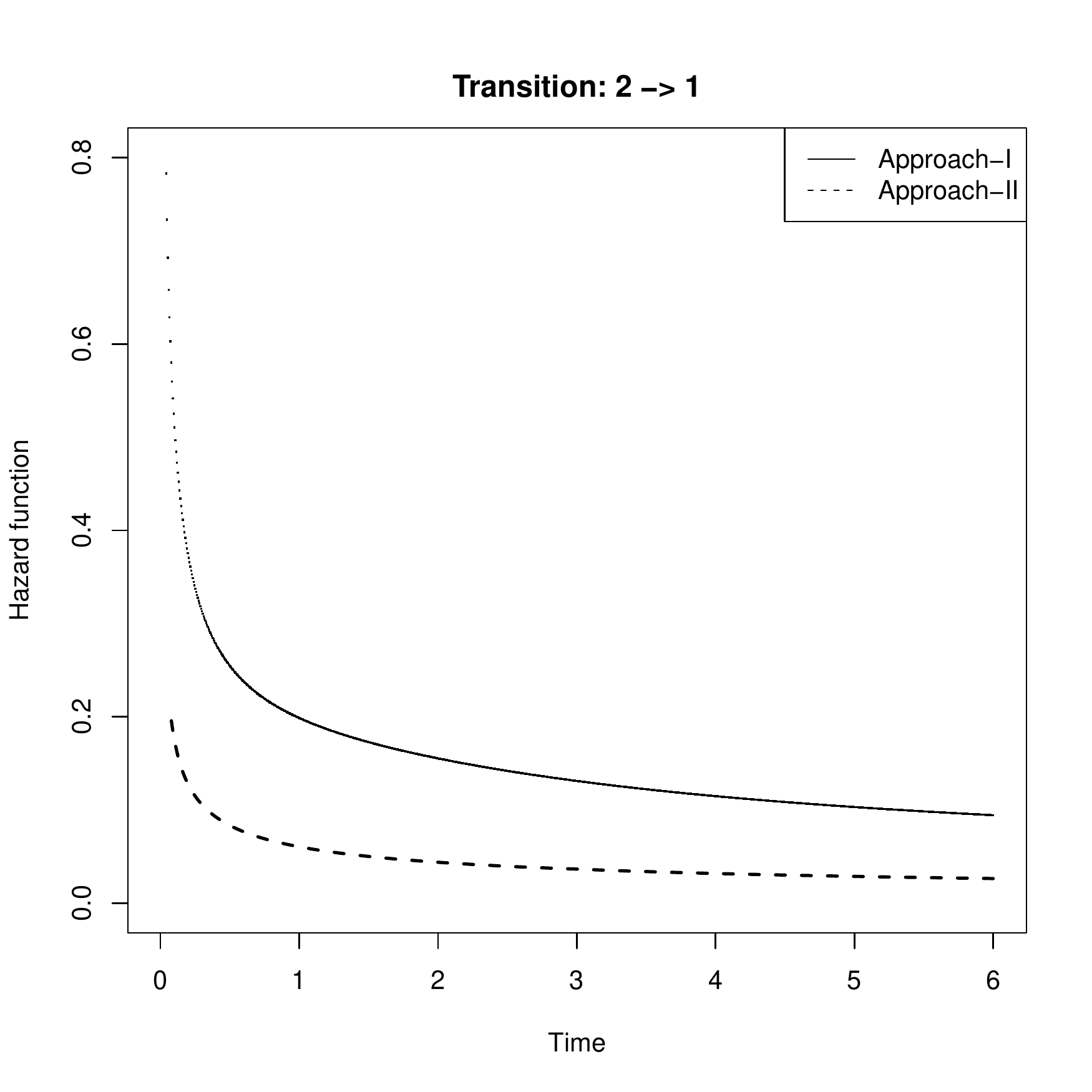}
  \includegraphics[scale=0.22]{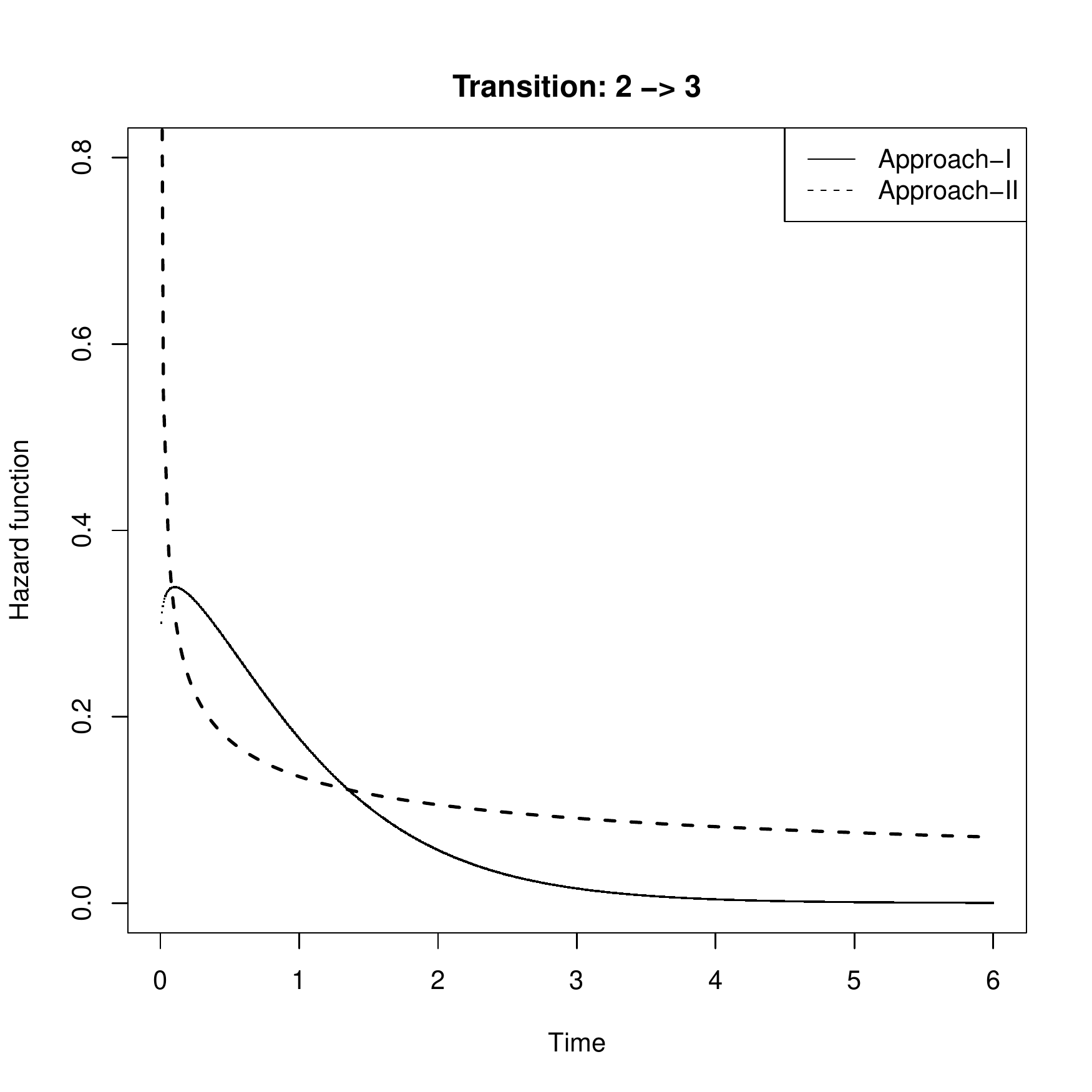}
 \caption{ \small Hazard rate of the semi-Markov process for each transition from the sojourn time and intensity transition approaches.}
\label{Fig:illustration2}
\end{figure}

For illustration purposes, we concentrate on a Weibull model as proposed in \cite{listwon2015semimarkov}. We perform a first full parametric proportional Weibull model for the sojourn times (Approach~I) with the BMI covariate for each transition. This yields significant results of BMI covariate for the transitions $1 \to 3$ and $3 \to 1$. Then, we decide to run a sparse model including BMI only for these transitions to facilitate the convergence of the model which can be difficult for a model with too many parameters as reported in \cite{listwon2015semimarkov}. For this new model, BMI regression coefficients remain significant for transitions $1 \to 3$ and $3 \to 1$ with $\widehat{\beta}_{13}=-0.88$ and $\widehat{\beta}_{31}=-0.448$, and respective $p$-values $0.012$, and $0.023$. The fact that hazard ratio of the sojourn time associated with these covariates is less than unity (estimated coefficients are negative), indicates that  BMI $\geq 25$ generally lengthens the duration of the sojourn time in state~1 when making a $1 \to 3$ transition and generally lengthens the duration of the sojourn time in state~3 when making a $3 \to 1$ transition. This can also be interpreted as a decrease of the risk of leaving ``optimal control'' state to ``unacceptable control'' as well as a decrease of the risk of leaving ``unacceptable control'' state to ``optimal control''. However, the magnitude of the estimated coefficients cannot be used to evaluate the change in the hazard ratio on the risk (recall equations \eqref{Eq:covariate-sojourn} and \eqref{Eq:covariate-approachII} and the differences between $\beta_{ij}$ and $\tilde{\beta}_{ij}$). We may also visualize the effect of BMI covariate on the associated risks by plotting the hazard rate of the semi-Markov model (intensity transition functions), deduced from the sojourn times. See Figure~\ref{Fig:illustration2} where we also plot estimated transition intensity functions using Approach~II estimation, which we describe now.

Approach~II presented in this paper offer a direct way to model the intensity transition functions and has the advantage of a split likelihood which facilitates an efficient optimization procedure. In the vignette available in \cite{Liq_github}, we present the results of the full Weibull proportional semi-Markov model using this Approach. For illustration, we do this in two ways: (i) Optimizing the full likelihood. (ii) Splitting the likelihood for each transition and optimizing individually. As expected, both ways yield the same likelihood  estimates. This highlight the advantage of the intensity transition based approach for overcoming potential computational issues in the optimization. Moreover, in some applications, some transitions of interest could then be further explored with any sophisticated survival model  of choice as the estimation for each transition is performed separately. See for example \cite{therneau2000cox,royston2002flexible,liquet2012investigating,jackson2016flexsurv}. 

We augment the plots in Figure \ref{Fig:illustration2} with the estimated transition intensity functions using Approach~II. Note that similarly to Approach~I, the BMI effects are in the same direction. However, the interpretation of the regression coefficients is different. For the intensity transition based estimation (Approach~II), the BMI regression coefficient estimates are $\hat{\tilde{\beta}}_{13} = -0.5$ and
 $\hat{\tilde{\beta}}_{31} = -0.67$. Here, in contrast to the Approach~I estimates, the exact magnitude of the coefficients may be interpreted as the change in the hazard ratios associated with  BMI $\geq 25$.

\section{Concluding Remarks}
\label{sec:conclusion}

We have surveyed the two main approaches for modeling and inference using semi-Markov processes in a parametric setting. These are Approach~I based on sojourn time distributions and the embedded Markov chain, and Approach~II based on transition intensity functions. Each approach has its advantages and drawbacks when carrying out inference and analysis and these were described in the paper. We further summarized the formulation of these two approaches and showed relations between them, where we focused on inference for each of the approaches. 

In general, the intensity transition based approach (Approach~II) allows the likelihood to be split when each transition has its own set of parameters. Such separation into two-state models facilitates more efficient optimization as well as usage of additional modeling tools. The large scale impact of such separation on bigger datasets and/or simulated data as part of a simulation study remains to be carried out in future work. However, even putting the computational aspects aside, there is great value in such separation as it allows to use many survival analysis R packages directly within the context of semi-Markov processes. 

Such packages include \texttt{flexsurv}  \cite{jackson2016flexsurv} which allows to fit various shapes of transition intensity functions including the Royston-Parmar spline model \cite{royston2002flexible}. Further, in a semi-parametric framework the \texttt{mstate} package \cite{de2011mstate} is popular for running multi-state models including the SMP. By exploiting the separation of the likelihood, \texttt{mstate} provides the estimation of covariate effects using Cox regression models. In addition, non-linear effects of the covariates could be investigated through smooth functions using the \texttt{mgcv} package which enables to run GAM models with survival data, see \cite{Rwood}. Further, with Approach~II, we can easily handle random effects using the \texttt{frailtypack} package  \cite{frailtyypack}, as has been exploited by \cite{liquet2012investigating}. Further, in presence of high dimensional data, elastic-net penalty \cite{coxnet} can be used via the package \texttt{glmnet} to fit Cox regression models. Such benefits of using the intensity transition based approach have previously been reported and exploited for general Markov multi-state models as in \cite{hougaard1999multi,andersen2002multi} and \cite{meira2009multi}. Our purpose here was to survey these benefits and compare to Approach~I as is popularly used with the { \texttt{SemiMarkov}}  package providing a flexible tool for inference of semi-Markov models based on sojourn time hazard rates in a parametric framework \cite{listwon2015semimarkov}.


{The focus in this paper was purely on time-homogenous SMP in which the absolute time or date is assumed not to affect the probabilistic behavior. In certain situations one should consider time-inhomogeneous models allowing the absolute calendar time to play a role in the probabilistic model and the inference. Such situations can occur if seasonal (periodic) phenomena are present, or in long term longitudinal studies where the nature of the cohort is assumed to vary during the study. In a future study, it may be of interest to contrast Approach~I and Approach~II in a time-inhomogeneous setting.}

Of further interest, we highlight the fact that PH (Phase Type) distributions may fit naturally in an SMP framework. Such approaches have been previously proposed in \cite{latouche1982phase}, \cite{malhotra1993selecting}, \cite{titman2010semi} and \cite{aalen1995phase}. Interestingly, as PH distributions provide a dense semi-parameteric approximation of any distribution of a non-negative random variable, it may be of interest to approximate semi-Markov processes with continuous time Markov chains constructed via PH distributions, where there are potentially more than $\ell$ states in the approximating Markov chain. To the best of our knowledge, such a semi-parametric inference setup has still not been explored.  Such a setup may work well with the EM algorithm for parameter estimation of this type of PH distributions, see \cite{asmussen1996fitting}. We leave this for future work.

\section*{Acknowledgements}
AA and BL are supported by the Australian Research Council (ARC) Centre of Excellence for Mathematical and Statistical Frontiers (ACEMS) under grant number CE140100049. YN is supported by Australian Research Council (ARC) under grant number DP180101602.

\vspace{-0.28cm}
\bibliography{Biblio}

\end{document}